\documentclass[paper,notoc]{JHEP3}
\usepackage{axodraw4j}
\usepackage{amsmath} 
\usepackage{graphicx}

\newcommand\new{\newcommand}         

\new{\Di}[1]{{\cal D}_{#1}}
\new{\Order}[1]{\mathcal{O}\(#1\)}
\new{\NLmap}[2]{#1 \mapsto \beta \(#1,#2\)}
\new{\U}[1]{\langle #1\rangle}
\new{\Ubar}[1]{[ #1 ]}
\new{\s}[1]{s_{#1} }

\def\beq{\begin{equation}}   
\def\eeq{\end{equation}}
\def\bea{\begin{eqnarray}}  
\def\eea{\end{eqnarray}} 
\def\nn{\nonumber}
\def\eps{\epsilon}

\def\P{\mathcal{P}}
\def\Q{\mathcal{Q}}
\def\A{\mathcal{A}}

\def\X{\textbf{X}}
\def\Y{\textbf{Y}}
\def\S{\textbf{S}}

\def\({\left(}
\def\){\right)}

\def\n3lo{$\mathrm{N^3LO}$}

\def\beq{\begin{equation}}
\def\eeq{\end{equation}}
\def\bsp#1\esp{\begin{split}#1\end{split}}

\newcommand{\ord}[0]{\mathcal{O}}

\newcommand{\XOne}[0]{\textbf{X}_1}
\newcommand{\XThree}[0]{\textbf{X}_2}
\newcommand{\XFour}[0]{\textbf{X}_3}
\newcommand{\XFive}[0]{\textbf{X}_4}
\newcommand{\XSix}[0]{\textbf{X}_5}
\newcommand{\XEight}[0]{\textbf{X}_6}
\newcommand{\XTen}[0]{\textbf{X}_{7}}
\newcommand{\XEleven}[0]{\textbf{X}_{8}}
\newcommand{\XTwelve}[0]{\textbf{X}_{9}}
\newcommand{\XThirteen}[0]{\textbf{X}_{10}}
\newcommand{\XNineNine}[0]{\textbf{X}_{11}}
\newcommand{\XFourteen}[0]{\textbf{X}_{11a}}
\newcommand{\XFifteen}[0]{\textbf{X}_{12}}
\newcommand{\XSixteen}[0]{\textbf{X}_{13}}
\newcommand{\XSeventeen}[0]{\textbf{X}_{14}}
\newcommand{\XEighteen}[0]{\textbf{X}_{15}}
\newcommand{\XTwenty}[0]{\textbf{X}_{16}}
\newcommand{\XTwentyFive}[0]{\textbf{X}_{17}}
\newcommand{\XThirty}[0]{\textbf{X}_{18}}

\newcommand{\SOne}[0]{\textbf{S}_1}
\newcommand{\SThree}[0]{\textbf{S}_2}
\newcommand{\SFour}[0]{\textbf{S}_3}
\newcommand{\SFive}[0]{\textbf{S}_4}
\newcommand{\SSix}[0]{\textbf{S}_5}
\newcommand{\SEight}[0]{\textbf{S}_6}
\newcommand{\STen}[0]{\textbf{S}_{7}}
\newcommand{\SEleven}[0]{\textbf{S}_{8}}
\newcommand{\STwelve}[0]{\textbf{S}_{9}}
\newcommand{\SThirteen}[0]{\textbf{S}_{10}}

\newcommand{\SFourteen}[0]{\textbf{S}_{11a}}
\newcommand{\SFifteen}[0]{\textbf{S}_{12}}
\newcommand{\SSixteen}[0]{\textbf{S}_{13}}
\newcommand{\SSeventeen}[0]{\textbf{S}_{14}}
\newcommand{\SEighteen}[0]{\textbf{S}_{15}}
\newcommand{\STwenty}[0]{\textbf{S}_{16}}
\newcommand{\STwentyFive}[0]{\textbf{S}_{17}}
\newcommand{\SThirty}[0]{\textbf{S}_{18}}

\def\xtri#1{
\begin{picture}(80, 60)(0, #1)
\Text(5,10)[]{$p_2$}
\Text(5,50)[]{$p_1$}
\ArrowLine(10,10)(20, 10)
\ArrowLine(10,50)(20,50)
\Line(20,10)(60,30)
\Line(20,50)(60,30)
\Line(20,50)(40,20)
\Line(20,10)(40,40)
\DashLine(70,40)(70,20){4}
\SetWidth{2}
\Line(60,30)(80,30)
\end{picture}
}

\def\tri#1{
\begin{picture}(80, 60)(0, #1)
\Text(5,10)[]{$p_2$}
\Text(5,50)[]{$p_1$}
\ArrowLine(10,10)(20, 10)
\ArrowLine(10,50)(20,50)
\Line(20,10)(60,30)
\Line(20,50)(60,30)
\DashLine(70,40)(70,20){4}
\Oval(20,30)(20,8)(0)
\SetWidth{2}
\Line(60,30)(80,30)
\end{picture}
}

\def\suns#1{
\begin{picture}(80, 60)(0, #1)
\Text(5,20)[]{$p_2$}
\Text(5,40)[]{$p_1$}
\ArrowLine(10,20)(20, 30)
\ArrowLine(10,40)(20,30)
\Line(20,30)(60,30)
\CArc(40,30)(20,0,360)
\DashLine(67,40)(67,20){4}
\SetWidth{2}
\Line(60,30)(74,30)
\end{picture}
}

\def\glass#1{
\begin{picture}(80, 60)(0, #1)
\Text(5,20)[]{$p_2$}
\Text(5,40)[]{$p_1$}
\ArrowLine(10,20)(20, 30)
\ArrowLine(10,40)(20,30)
\CArc(30,30)(10,0,360)
\CArc(50,30)(10,0,360)
\DashLine(70,40)(70,20){4}
\SetWidth{2}
\Line(60,30)(80,30)
\end{picture}
}

\def\cglass#1{
\begin{picture}(90, 60)(0, #1)
\Text(5,20)[]{$p_2$}
\Text(5,40)[]{$p_1$}
\Text(87,20)[]{$p_2$}
\Text(87,40)[]{$p_1$}
\ArrowLine(70,30)(80, 40)
\ArrowLine(70,30)(80,20)
\ArrowLine(10,20)(20, 30)
\ArrowLine(10,40)(20,30)
\CArc(30,30)(10,0,360)
\CArc(60,30)(10,0,360)
\DashLine(45,40)(45,20){4}
\SetWidth{2}
\Line(40,30)(50,30)
\end{picture}
}

\def\YA#1{
\begin{picture}(80,60)(0,#1)
\Text(5,20)[]{$p_2$}
\Text(5,40)[]{$p_1$}
\Text(73,10)[]{$p_2$}
\Text(73,50)[]{$p_1$}
\ArrowLine(10,20)(20, 30)
\ArrowLine(10,40)(20,30)
\ArrowLine(55,50)(65,50)
\ArrowLine(55,10)(65,10)
\Line(20,30)(55,10)
\Oval(55,30)(20,5)(0)
\DashLine(37.5,50)(37.5,10){4}
\SetWidth{2}
\Line(20, 30)(55, 50)
\end{picture}
}

\def\YB#1{
\begin{picture}(80,60)(0,#1)
\Text(5,20)[]{$p_2$}
\Text(5,40)[]{$p_1$}
\Text(73,20)[]{$p_2$}
\Text(73,40)[]{$p_1$}
\ArrowLine(10,20)(20, 30)
\ArrowLine(10,40)(20,30)
\ArrowLine(55,30)(65,40)
\ArrowLine(55,30)(65,20)
\CArc(37.5, 30)(17.5, 180, 450)
\CArc(55,47.5)(17.5,180,270)
\DashLine(32.5,50)(32.5,10){4}
\SetWidth{2}
\CArc(37.5, 30)(17.5, 90, 180)
\end{picture}
}

\def\YC#1{
\begin{picture}(80,60)(0,#1)
\Text(5,20)[]{$p_2$}
\Text(5,40)[]{$p_1$}
\Text(73,10)[]{$p_2$}
\Text(73,50)[]{$p_1$}
\ArrowLine(10,20)(20, 30)
\ArrowLine(10,40)(20,30)
\ArrowLine(55,50)(65,50)
\ArrowLine(55,10)(65,10)
\Line(20,30)(55,10)
\Line(45,44.28)(55,50)
\Line(45,44.28)(55,10)
\Line(55,50)(55,10)
\DashLine(37.5,50)(37.5,10){4}
\SetWidth{2}
\Line(20, 30)(45, 44.28)
\end{picture}
}

\def\YE#1{
\begin{picture}(80, 60)(0, #1)
\Text(5,20)[]{$p_2$}
\Text(5,40)[]{$p_1$}
\Text(75,20)[]{$p_2$}
\Text(75,40)[]{$p_1$}
\ArrowLine(10,20)(20, 30)
\ArrowLine(10,40)(20,30)
\ArrowLine(60,30)(70,40)
\ArrowLine(60,30)(70,20)
\CArc(30,30)(10,180,360)
\CArc(50,30)(10,0,360)
\DashLine(30,50)(30,10){4}
\SetWidth{2}
\CArc(30,30)(10,0,180)
\end{picture}
}

\def\YD#1{
\begin{picture}(80,60)(0,#1)
\Text(5,10)[]{$p_2$}
\Text(5,50)[]{$p_1$}
\Text(73,10)[]{$p_1$}
\Text(73,50)[]{$p_2$}
\ArrowLine(10,50)(20, 50)
\ArrowLine(10,10)(20,10)
\ArrowLine(55,50)(65,50)
\ArrowLine(55,10)(65,10)
\Line(20,10)(55,10)
\Line(55,50)(55,10)
\Line(37.5,50)(37.5,10)
\Line(20,50)(20,10)
\Line(35, 50)(55, 50)
\DashLine(28.25,55)(28.25,5){4}
\SetWidth{2}
\Line(20, 50)(37.5, 50)
\end{picture}
}

\def\YF#1{
\begin{picture}(80,60)(0,#1)
\Text(5,20)[]{$p_2$}
\Text(5,40)[]{$p_1$}
\Text(73,10)[]{$p_2$}
\Text(73,50)[]{$p_1$}
\ArrowLine(10,20)(20, 30)
\ArrowLine(10,40)(20,30)
\ArrowLine(55,50)(65,50)
\ArrowLine(55,10)(65,10)
\Line(20,30)(55,10)
\Line(45,44.28)(55,50)
\Line(45,44.28)(55,10)
\Line(55,50)(45,16.72)
\DashLine(37.5,50)(37.5,10){4}
\SetWidth{2}
\Line(20, 30)(45, 44.28)
\end{picture}
}

\def\Xone#1{
\begin{picture}(80,60)(0,#1)
\Text(5,20)[]{$p_2$}
\Text(5,40)[]{$p_1$}
\Text(77,20)[]{$p_2$}
\Text(77,40)[]{$p_1$}
\ArrowLine(10,20)(20, 30)
\ArrowLine(10,40)(20,30)
\ArrowLine(60,30)(70,40)
\ArrowLine(60,30)(70,20)
\CArc(40,30)(20,0,360)
\DashLine(40,55)(40,5){4}
\SetWidth{2}
\Line(20, 30)(60, 30)
\end{picture}
}

\def\Xeleven#1{
\begin{picture}(80,60)(0,#1)
\Text(5,10)[]{$p_2$}
\Text(5,50)[]{$p_1$}
\Text(73,10)[]{$p_1$}
\Text(73,50)[]{$p_2$}
\ArrowLine(10,50)(20, 50)
\ArrowLine(10,10)(20,10)
\ArrowLine(55,50)(65,50)
\ArrowLine(55,10)(65,10)
\Line(20,10)(55,10)
\Line(55,50)(55,10)
\Line(20,50)(20,10)
\Line(35, 50)(55, 50)
\DashLine(26,55)(49,5){4}
\Line(20, 50)(37.5, 50)
\SetWidth{2}
\Line(37.5,50)(37.5,10)
\end{picture}
}

\def\Xfourteen#1{
\begin{picture}(80,60)(0,#1)
\Text(5,20)[]{$p_2$}
\Text(5,40)[]{$p_1$}
\Text(73,20)[]{$p_2$}
\Text(73,40)[]{$p_1$}
\ArrowLine(10,20)(20, 30)
\ArrowLine(10,40)(20,30)
\ArrowLine(55,30)(65,40)
\ArrowLine(55,30)(65,20)
\DashCArc(37.5, 30)(17.5, 90, 180){1.5}
\CArc(37.5, 30)(17.5, 360, 450)
\CArc(55,47.5)(17.5,180,270)
\DashLine(44.5,50)(44.5,10){4}
\SetWidth{2}
\CArc(37.5, 30)(17.5, 180, 360)
\end{picture}
}

\def\Xten#1{
\begin{picture}(80,60)(0,#1)
\Text(5,20)[]{$p_2$}
\Text(5,40)[]{$p_1$}
\Text(73,20)[]{$p_2$}
\Text(73,40)[]{$p_1$}
\ArrowLine(10,20)(20, 30)
\ArrowLine(10,40)(20,30)
\ArrowLine(55,30)(65,40)
\ArrowLine(55,30)(65,20)
\CArc(37.5, 30)(17.5, 0, 360)
\DashLine(30,50)(45,10){4}
\SetWidth{2}
\Line(37.5, 47.50)(37.5,12.50)
\end{picture}
}

\def\Xthirteen#1{
\begin{picture}(80,60)(0,#1)
\Text(5,10)[]{$p_2$}
\Text(5,50)[]{$p_1$}
\Text(73,10)[]{$p_2$}
\Text(73,50)[]{$p_1$}
\ArrowLine(10,50)(20, 50)
\ArrowLine(10,10)(20,10)
\ArrowLine(55,50)(65,50)
\ArrowLine(55,10)(65,10)
\Line(20,10)(55,10)
\Line(55,50)(55,10)
\Line(20,50)(20,10)
\Line(35, 50)(55, 50)
\DashLine(26,55)(49,5){4}
\Line(20, 50)(37.5, 50)
\SetWidth{2}
\Line(37.5,50)(37.5,10)
\end{picture}
}

\def\Xtwentyfive#1{
\begin{picture}(80,60)(0,#1)
\Text(5,10)[]{$p_2$}
\Text(5,50)[]{$p_1$}
\Text(73,10)[]{$p_1$}
\Text(73,50)[]{$p_2$}
\ArrowLine(10,50)(20, 50)
\ArrowLine(10,10)(20,10)
\ArrowLine(55,50)(65,50)
\ArrowLine(55,10)(65,10)
\Line(20,10)(55,10)
\Line(55,50)(55,10)
\Line(37.5,50)(37.5,10)
\Line(20,50)(20,10)
\Line(35, 50)(55, 50)
\DashLine(26,55)(49,5){4}
\Line(20,10)(55,10)
\SetWidth{2}
\Line(20, 50)(37.5, 50)
\end{picture}
}
\def\Xfive#1{
\begin{picture}(80,60)(0,#1)
\Text(5,10)[]{$p_2$}
\Text(5,50)[]{$p_1$}
\Text(73,10)[]{$p_1$}
\Text(73,50)[]{$p_2$}
\ArrowLine(10,50)(20, 50)
\ArrowLine(10,10)(20,10)
\ArrowLine(55,50)(65,50)
\ArrowLine(55,10)(65,10)
\Line(20,10)(55,10)
\Line(55,50)(55,10)
\Line(20,30)(55,30)
\Line(20,50)(20,10)
\DashLine(37.5,55)(37.5,5){4}
\Line(20,10)(55,10)
\SetWidth{2}
\Line(20, 50)(55, 50)
\end{picture}
}
\def\Xeight#1{
\begin{picture}(80,60)(0,#1)
\Text(5,10)[]{$p_2$}
\Text(5,50)[]{$p_1$}
\Text(73,10)[]{$p_1$}
\Text(73,50)[]{$p_2$}
\ArrowLine(10,50)(20, 50)
\ArrowLine(10,10)(20,10)
\ArrowLine(55,50)(65,50)
\ArrowLine(55,10)(65,10)
\Line(20,10)(55,10)
\Line(55,50)(55,10)
\Line(20,50)(20,10)
\DashLine(37.5,55)(37.5,5){4}
\Line(20,10)(55,10)
\Line(20, 50)(55, 50)
\SetWidth{2}
\Line(20,30)(55,30)
\end{picture}
}

\def\Xthree#1{
\begin{picture}(80,60)(0,#1)
\Text(5,10)[]{$p_2$}
\Text(5,50)[]{$p_1$}
\Text(73,10)[]{$p_1$}
\Text(73,50)[]{$p_2$}
\ArrowLine(10,50)(20, 50)
\ArrowLine(10,10)(20,10)
\ArrowLine(55,50)(65,50)
\ArrowLine(55,10)(65,10)
\Line(55,50)(55,10)
\Line(20,50)(20,10)
\DashLine(37.5,55)(37.5,0){3.5}
\Oval(37.5,10)(4,17.5)(0)
\SetWidth{2}
\Line(20, 50)(55, 50)
\end{picture}
}

\def\Xsix#1{
\begin{picture}(80,60)(0,#1)
\Text(5,10)[]{$p_2$}
\Text(5,50)[]{$p_1$}
\Text(73,10)[]{$p_1$}
\Text(73,50)[]{$p_2$}
\ArrowLine(10,50)(20, 50)
\ArrowLine(10,10)(20,10)
\ArrowLine(55,50)(65,50)
\ArrowLine(55,10)(65,10)
\Line(55,50)(55,10)
\Line(20,50)(20,10)
\DashLine(40,55)(40,0){3.5}
\Line(20, 28)(55, 10)
\Line(20, 10)(55, 32)
\SetWidth{2}
\Line(20, 50)(55, 50)
\end{picture}
}

\def\Xthirty#1{
\begin{picture}(80,60)(0,#1)
\Text(5,20)[]{$p_2$}
\Text(5,40)[]{$p_1$}
\Text(73,10)[]{$p_2$}
\Text(73,50)[]{$p_1$}
\ArrowLine(10,20)(20, 30)
\ArrowLine(10,40)(20,30)
\ArrowLine(55,50)(65,50)
\ArrowLine(55,10)(65,10)
\Line(20,30)(55,10)
\Line(45,44.28)(55,50)
\Line(45,44.28)(55,10)
\Line(55,50)(45,16.72)
\DashLine(37,55)(52,5){4}
\SetWidth{2}
\Line(20, 30)(45, 44.28)
\end{picture}
}

\def\Xeighteen#1{
\begin{picture}(80,60)(0,#1)
\Text(5,20)[]{$p_2$}
\Text(5,40)[]{$p_1$}
\Text(73,10)[]{$p_1$}
\Text(73,50)[]{$p_2$}
\ArrowLine(10,20)(20, 30)
\ArrowLine(10,40)(20,30)
\ArrowLine(55,50)(65,50)
\ArrowLine(55,10)(65,10)
\Line(20,30)(55,10)
\Line(45,44.28)(55,50)
\Line(45,44.28)(55,10)
\Line(20, 30)(45, 44.28)
\DashLine(37,55)(52,5){4}
\SetWidth{2}
\Line(55,50)(45,16.72)
\end{picture}
}

\def\Xtwenty#1{
\begin{picture}(80,60)(0,#1)
\Text(5,10)[]{$p_2$}
\Text(5,50)[]{$p_1$}
\Text(73,10)[]{$p_2$}
\Text(73,50)[]{$p_1$}
\ArrowLine(10,10)(20, 10)
\ArrowLine(10,50)(20,50)
\ArrowLine(55,50)(65,50)
\ArrowLine(55,10)(65,10)
\Line(20,10)(55,10)
\Line(40,50)(55,10)
\Line(20, 50)(55, 50)
\Line(20, 50)(20,10)
\DashLine(32,55)(51,5){4}
\SetWidth{2}
\Line(55,50)(40,10)
\end{picture}
}

\def\Xseventeen#1{
\begin{picture}(80,60)(0,#1)
\Text(5,20)[]{$p_2$}
\Text(5,40)[]{$p_1$}
\Text(73,10)[]{$p_1$}
\Text(73,50)[]{$p_2$}
\ArrowLine(10,20)(20, 30)
\ArrowLine(10,40)(20,30)
\ArrowLine(55,50)(65,50)
\ArrowLine(55,10)(65,10)
\Line(20,30)(55,10)
\Line(45,44.28)(55,50)
\Line(20, 30)(45, 44.28)
\Line(55,50)(55,10)
\DashLine(37,55)(52,5){4}
\SetWidth{2}
\Line(55,50)(37.5,20)
\end{picture}
}

\def\Xfour#1{
\begin{picture}(80,60)(0,#1)
\Text(5,20)[]{$p_2$}
\Text(5,40)[]{$p_1$}
\Text(73,10)[]{$p_2$}
\Text(73,50)[]{$p_1$}
\ArrowLine(10,20)(20, 30)
\ArrowLine(10,40)(20,30)
\ArrowLine(55,50)(65,50)
\ArrowLine(55,10)(65,10)
\Line(20,30)(55,10)
\Line(45,44.28)(55,50)
\Line(20, 30)(45, 44.28)
\Line(55,50)(55,10)
\DashLine(37,55)(52,5){4}
\SetWidth{2}
\Line(20,30)(55,30)
\end{picture}
}

\def\Xsixteen#1{
\begin{picture}(80,60)(0,#1)
\Text(5,20)[]{$p_2$}
\Text(5,40)[]{$p_1$}
\Text(73,10)[]{$p_1$}
\Text(73,50)[]{$p_2$}
\ArrowLine(10,20)(20, 30)
\ArrowLine(10,40)(20,30)
\ArrowLine(55,50)(65,50)
\ArrowLine(55,10)(65,10)
\Line(20,30)(55,50)
\Line(20,30)(37.5,20)
\Line(20, 30)(45, 44.28)
\Line(55,50)(55,10)
\Line(55,50)(37.5,20)
\DashLine(37,55)(52,5){4}
\SetWidth{2}
\Line(37.5, 20)(55,10)
\end{picture}
}

\def\Xfifteen#1{
\begin{picture}(80,60)(0,#1)
\Text(5,10)[]{$p_2$}
\Text(5,50)[]{$p_1$}
\Text(73,10)[]{$p_2$}
\Text(73,50)[]{$p_1$}
\ArrowLine(10,10)(20, 10)
\ArrowLine(10,50)(20,50)
\ArrowLine(55,50)(65,50)
\ArrowLine(55,10)(65,10)
\Line(20,10)(55,10)
\Line(20,50)(20,10)
\Line(20, 50)(55, 50)
\DashLine(55, 50)(55,10){2}
\DashLine(32,55)(51,5){4}
\Line(20,10)(45,50)
\SetWidth{2}
\Line(20,50)(45,50)
\end{picture}
}

\def\Xtwelve#1{
\begin{picture}(80,60)(0,#1)
\Text(5,10)[]{$p_2$}
\Text(5,50)[]{$p_1$}
\Text(73,10)[]{$p_2$}
\Text(73,50)[]{$p_1$}
\ArrowLine(10,10)(20, 10)
\ArrowLine(10,50)(20,50)
\ArrowLine(55,50)(65,50)
\ArrowLine(55,10)(65,10)
\Line(20,10)(55,10)
\Line(20,50)(20,10)
\Line(20, 50)(55, 50)
\Line(55, 50)(55,10)
\DashLine(32,55)(51,5){4}
\SetWidth{2}
\Line(20,10)(55,50)
\end{picture}
}

\allowdisplaybreaks[1]

\title{NNLO phase space master integrals for two-to-one inclusive cross sections in dimensional regularization}

\author{Charalampos Anastasiou\\
  Institute for Theoretical Physics, ETH Zurich,
  8093 Zurich, Switzerland\\
  E-mail: \email{babis@phys.ethz.ch}}

\author{Stephan Buehler\\
  Institute for Theoretical Physics, ETH Zurich,
  8093 Zurich, Switzerland\\
  E-mail: \email{buehler@itp.phys.ethz.ch}}

\author{Claude Duhr\\
  Institute for Theoretical Physics, ETH Zurich,
  8093 Zurich, Switzerland\\
  E-mail: \email{duhrc@itp.phys.ethz.ch}}

\author{Franz Herzog\\
  Institute for Theoretical Physics, ETH Zurich,
  8093 Zurich, Switzerland\\
  E-mail: \email{fherzog@itp.phys.ethz.ch}}

\abstract{
We evaluate all phase space master integrals which are required for the total cross section of generic $2 \to 1$ processes at NNLO
as a series expansion in the dimensional regulator $\epsilon$. Away from the limit of threshold production, our expansion includes one order higher than what has 
been available in the literature. At threshold, we provide expressions which are valid to all orders in terms of $\Gamma$ functions and hypergeometric 
functions.  These results are a necessary ingredient for the renormalization and mass factorization of singularities in $2 \to 1$ inclusive cross sections 
at \n3lo in QCD.
 }


\preprint{}

\begin{document}

\catcode`\@=11
\font\manfnt=manfnt
\def\Watchout{\@ifnextchar [{\W@tchout}{\W@tchout[1]}}
\def\W@tchout[#1]{{\manfnt\@tempcnta#1\relax%
  \@whilenum\@tempcnta>\z@\do{%
    \char"7F\hskip 0.3em\advance\@tempcnta\m@ne}}}
\let\foo\W@tchout
\def\dubious{\@ifnextchar[{\@dubious}{\@dubious[1]}}
\let\enddubious\endlist
\def\@dubious[#1]{%
  \setbox\@tempboxa\hbox{\@W@tchout#1}
  \@tempdima\wd\@tempboxa
  \list{}{\leftmargin\@tempdima}\item[\hbox to 0pt{\hss\@W@tchout#1}]}
\def\@W@tchout#1{\W@tchout[#1]}
\catcode`\@=12


\section{Introduction}
\label{sec:introduction}

Signals of novel physics, such as the recently discovered Higgs boson 
by the ATLAS~\cite{:2012gk} and CMS~\cite{:2012gu} collaborations, as well as  standard candles, 
such as Drell-Yan production of electroweak 
gauge bosons~\cite{Chatrchyan:2011wt,Aad:2011qv}, have been at the epicenter of experiments at hadron colliders. 
The analysis and interpretation of data relies crucially on high precision theoretical estimates, based on perturbation theory, for the rates of $2 \to 1$ processes. 
Most importantly, higher order corrections due to Quantum Chromodynamics (QCD) are generally large at LHC energies, altering the central value and lowering
the theoretical uncertainty of many observables.

The state of the art in fixed-order perturbative QCD for $2 \to 1$ processes is next-to-next-to-leading-order (NNLO). 
The first inclusive cross section computation at this order was performed more than two decades ago for Drell-Yan production~\cite{Hamberg:1990np}.
This calculation was followed by several computations for inclusive Higgs production cross sections~\cite{Harlander:2002wh,Anastasiou:2002yz,Anastasiou:2002wq,Harlander:2002vv,Harlander:2003ai,Brein:2003wg,Pak:2009dg,Harlander:2009mq,Pak:2011hs,Ravindran:2003um} with the emergence of powerful computational 
techniques~\cite{Harlander:2002wh,Anastasiou:2002yz}.    

The precision of NNLO cross sections is quite high as indicated by the study scale variations.  
For example, the uncertainty due to scale variations in Higgs production via gluon fusion is less than $10\%$ and in Drell-Yan production 
less than $2\%$ (see, for example, ref.~\cite{Anastasiou:2012hx} and ref.~\cite{Anastasiou:2003ds}). 
However, a more reliable estimate of the uncertainty due to the truncation of perturbation theory would be the evaluation of yet another term in 
the series expansion.  
Now that a Higgs boson has been discovered, measuring its properties and couplings  precisely becomes a priority. 
With this prospect of precision physics in the Higgs sector and the paramount importance of Higgs coupling measurements for our understanding of physics at high energies, 
it is evident that many important observables will need to be validated at an even higher order in perturbation theory, i.e. at next-to-next-to-next-to-leading order (NNNLO or \n3lo).

In this paper, we present one of the ingredients needed for $2 \to 1 $ inclusive cross sections at \n3lo. In particular, we compute the double-real and real-virtual
master integrals for a generic $2\to1$ scattering amplitude at one order higher in the dimensional regularization parameter $\epsilon=(4-D)/2$, with $D$ the space-time dimension, than previously known~\cite{Anastasiou:2002yz}. The paper is organized as follows. In section \ref{sec:NNNLO}, we motivate our computation by explaining where they enter the calculation of
a \n3lo observable. In section \ref{sec:method} we provide some details about the method we use.
Sections \ref{sec:RR} and \ref{sec:RV} contain the results and we conclude in section \ref{sec:Conclusions}.
\section{NNLO contributions to N$^\mathbf{3}$LO observables}
\label{sec:NNNLO}

In this paper we are concerned with the computation of the fully inclusive cross section for the production of a massive state $X$.
In perturbation theory this cross section can schematically be written as
\beq
\sigma_X = \tau \sum_{i,j}f^{(0)}_i\otimes f^{(0)}_j\otimes \sum_{n=0}^\infty\alpha_s^{(0)n}\frac{\hat{\sigma}_{ij,n}(z)}{z}\,,
\eeq
where 
\beq
\tau=\frac{m_X^2}{S} \qquad \text{and} \qquad z=\frac{m_X^2}{s},
\eeq
with $S$ being the total center of mass energy, $s$ the partonic center of mass energy carried by the momenta of the incoming partons, $p_1$ and $p_2$,
 and $m_X$ the (on-shell) mass of particle $X$. The bare strong coupling constant is denoted as $\alpha_s^{(0)}$, $f^{(0)}_i$ are the bare parton distribution functions (PDF's) and 
$\hat{\sigma}_{ij,n}$ is the partonic  cross section for $X+n\textrm{ jets}$, which in turn admits a perturbative expansion in the number of loops. 
$\hat{\sigma}_{ij,n}$ may be written as a phase space integral over the squared $X+n\textrm{ jets}$ amplitude as follows,
\beq
\hat{\sigma}_{ij,n}=\frac{1}{2s}\int d\Phi_{n+1} |\A_{ij\to X+n\textrm{ jets}}|^2 \, ,
\eeq
where we define the measure of the phase space volume for a massive particle $X$ of mass $m_X$ and $n-1$ massless particles by
\beq
d\Phi_n=     \frac{d^Dp_X}{(2\pi)^{D-1}} \delta_{+}(p_X^2-m_X^2)  \(\prod_{i=3}^{n+1} \frac{d^Dp_i}{(2\pi)^{D-1}}\delta_{+}(p_i^2)\)  (2\pi)^D
\delta^{(D)}(p_{1...n+1}-p_X) \,,
\eeq
where 
\beq
p_{i_1...i_n} = \tau_{i_1}p_{i_1}+...+\tau_{i_n}p_{i_n}\,,\qquad \tau_i = 
\left\{ 
\begin{array}{ll} 
+1      &\quad\mbox{if  $i = 1,2$}\,,\\
-1      &\quad\mbox{if  $i > 2$}\,.
\end{array} 
\right. 
\eeq
Using this notation, the Lorentz invariants appearing in the phase space integrals are defined through 
\beq
s_{i_1..i_n}=(p_{i_1..i_n})^2\,.
\eeq

While the inclusive cross section $\sigma_X$ must obviously be finite order by order in perturbation theory, the individual pieces contributing to a given loop order are divergent. 
Final state infrared (IR) divergences cancel mutually between the real and virtual corrections, whereas ultraviolet (UV) and initial-state IR divergences have to be dealt with
by replacing the bare coupling and PDF's by their renormalized counterparts, which requires the introduction of explicit counterterms proportional to poles in the dimensional 
regularization parameter $\eps$ multiplying lower order coefficients of the cross sections. As an example, if $\hat{\sigma}^{(\ell)}$ denotes the $\ell$-th order correction
 to the cross section, the \n3lo UV/PDF counterterms can schematically be written as
\beq
\delta\hat{\sigma}^{(3)} \sim \frac{1}{\eps} C_1 \times \hat{\sigma}^{(2)} + \( \frac{1}{\eps^2} C_2  + \frac{1}{\eps} C_3 \)  \times \hat{\sigma}^{(1)}
+ \( \frac{1}{\eps^3} C_4  + \frac{1}{\eps^2} C_5+ \frac{1}{\eps} C_6\)  \times \hat{\sigma}^{(0)}\,,
\eeq
where the $\times$ may indicate a convolution in the case of the PDF counterterms. As a consequence, the poles of the counterterms produce finite contributions
 to the \n3lo cross section from the higher orders in the $\epsilon$ expansion of $\hat\sigma^{(\ell)}$. 

The aim of this paper is to provide the NNLO master integrals which are required in $\hat{\sigma}^{(2)}$ to one order higher in the $\epsilon$ expansion.
At this order in perturbation theory three different contributions need to be taken into account. As an example, the NNLO correction to the partonic cross section for $gg\to H$
gets contributions from the following three types of interference diagrams,
\begin{itemize}
\item  {\bf double-virtual:}  the two-loop production amplitude interfered with the Born amplitude as well as the square of the one-loop amplitude 
       for $gg \rightarrow H$, e.g.,
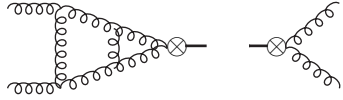
\begin{center}
\begin{picture}(250,37)(0,0)
\Gluon(40,5)(60,5){2}{4}
\Gluon(60,5)(60,35){2}{6}
\Gluon(40,35)(60,35){2}{4}
\Gluon(60,35)(100, 20){2}{8}
\Gluon(60,5)(100, 20){2}{8}
\Gluon(80,27.5)(80,12.5){2}{2}
\put(100,18){{$\otimes$}}
\SetWidth{1}
\Line(107,21)(115,21)
\Line(131,21)(138,21)
\SetWidth{0.5}
\put(134,18){{ $\otimes$}}
\Gluon(145,20)(165, 37){2}{4}
\Gluon(145,20)(165, 5){2}{4}
\end{picture}
\end{center}
\item {\bf real-virtual:} the one-loop production amplitudes for $gg \rightarrow Hg$, $gq \rightarrow Hq$ and 
  $g\bar q \rightarrow H \bar q$ interfered with the corresponding Born amplitudes, e.g.,
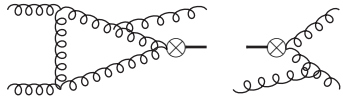
\begin{center}
\begin{picture}(250,40)(0,0)
\Gluon(40,5)(60,5){2}{4}
\Gluon(60,5)(60,35){2}{6}
\Gluon(40,35)(60,35){2}{4}
\Gluon(60,35)(100, 20){2}{8}
\Gluon(60,5)(100, 20){2}{8}
\Gluon(80,27.5)(115,35){2}{5}
\put(96,18){{ $\otimes$}}
\SetWidth{1}
\Line(107,21)(115,21)
\Line(130,21)(138,21)
\SetWidth{0.5}
\put(134,18){{ $\otimes$}}
\Gluon(145,20)(165, 35){2}{4}
\Gluon(145,20)(165, 5){2}{4}
\Gluon(157.5,12.5)(125,5){2}{5}
\end{picture}
\end{center}
\item  {\bf double-real:} the square of the Born amplitudes for 
  $gg \rightarrow Hgg$, 
  $gg \rightarrow Hq \bar q$, $gq \rightarrow Hgq$, $g\bar q \rightarrow Hg
  \bar q$, $qq \rightarrow Hqq$, and $q \bar q \rightarrow H q \bar q$, e.g., 
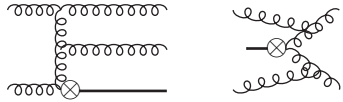
\begin{center}
\begin{picture}(250,37)(0,0)
\Gluon(40,5)(60,5){2}{4}
\Gluon(60,5)(60,35){2}{6}
\Gluon(40,35)(60,35){2}{4}
\Gluon(60,35)(100, 35){2}{8}
\Gluon(60,20)(100, 20){2}{8}
\put(56,2){{ $\otimes$}}
\SetWidth{1}
\Line(67,4.5)(100,4.5)
\Line(130,20.5)(138,20.5)
\SetWidth{0.5}
\put(134,18){{ $\otimes$}}
\Gluon(145,20)(165, 35){2}{4}
\Gluon(145,20)(165, 5){2}{4}
\Gluon(157.5,12.5)(125,5){2}{5}
\Gluon(157.5,27.5)(125,35){2}{5}
\end{picture}
\end{center}
\end{itemize}
 First, the double-virtual corrections decouple from the phase space integration, reducing the computation of the two-loop phase space
 integrals effectively to the computation of the two-loop QCD form factor. The master integrals for the two-loop QCD form factor have been evaluated in
 ref.~\cite{Gonsalves:1983nq,Kramer:1986sr,Gehrmann:2005pd} to all orders in $\eps$ in terms of $\Gamma$ functions and hypergeometric functions, which can easily
 be expanded in $\eps$ using the {\tt HypExp} package~\cite{Huber:2005yg}. For this reason we will not consider these integrals any further, but we simply list them
 in appendix~\ref{sec:Results_VV} for completeness.
Next, in ref.~\cite{Anastasiou:2002yz} it was shown how to compute the phase space integrals describing the real-virtual and double-real corrections as a Laurent expansion in the
 dimensional regulator up to order $\eps^0$. More precisely, it was shown that all the relevant integrals can be reduced to a small set of master integrals, which were
 computed to the order in $\eps$ required for NNLO computations. In the rest of this paper we compute all the master integrals to one order higher in the $\eps$ 
expansion, thus preparing the ground for using them in computations beyond NNLO.


\section{Phase space integrals from differential equations}
\label{sec:method}

In this section we give a brief account on how to compute the master integrals for the real-virtual and double-real phase space integrals.
Our computations are based on the method introduced in ref.~\cite{Anastasiou:2002yz}, which we shortly review in the following. 
In ref.~\cite{Anastasiou:2002yz} all real-virtual and double-real topologies were expressed as a two-loop forward scattering amplitude with,
 respectively, two and three cut propagators. Such cut integrals are related to the desired phase space integrals via Cutkosky's rule, 
\beq\label{eq:disc}
\textrm{Disc}\frac{1}{p^2-m^2+i\varepsilon} = 2\pi i\,\delta_+(p^2-m^2) = 2\pi i\,\delta(p^2-m^2)\,\theta(p^0)\,.
\eeq
Eq.~\eqref{eq:disc} allows us to interpret a phase space integral as a loop integral, and thus we can apply results  developed for the computation 
of Feynman integrals. In particular, these two-loop integrals can be reduced to a set of master
integrals by virtue of the Laporta algorithm~\cite{Laporta:2001dd}, using integration-by-parts (IBP) and Lorentz invariance identities,
as implemented in the program AIR~\cite{Anastasiou:2004vj}, with the additional constraint that integrals with some of the cut propagators not present vanish
 and need not to be considered in the reduction process~\cite{Anastasiou:2002yz}.

The master integrals themselves are calculated using the method of differential equations~\cite{Kotikov:1990kg,Gehrmann:1999as}: 
we differentiate each master integral under the integration sign with respect to a kinematic invariant, in our case the mass $m_X^2$ 
or equivalently the variable $z$.

During the differentiation process, loop integrals with higher powers of propagators are produced. They can again be expressed in terms of the master integrals
themselves via the IBP identities and the system of differential equations closes upon itself. If we denote the master integrals by $\textbf{F}_i(z,\eps)$, we arrive
at a set of first order ordinary linear differential equations for the master integrals,
\beq
\frac{\partial}{\partial z}\textbf{F}_i(z,\eps) = \sum_jc_{ij}(z,\eps)\,\textbf{F}_j(z,\eps)\,,
\eeq
where the $c_{ij}(z,\eps)$ are rational functions of $z$ and $\eps$ which have poles in $z$ at most at $z=0$ and/or $z=\pm 1$. In the case where the set 
of differential equations is triangular order by order in $\eps$,
the set of equations may be solved order by order using standard techniques for ordinary differential equations. In particular, if $\omega(z)$ is a 
homogeneous solution to the linear differential equation
\beq
\frac{\partial}{\partial z}y(z) = A(z)y(z)+B(z)\,,
\eeq
then the solution to the inhomogeneous differential equation is given by the integral
\beq\label{eq:solution}
y(z) = \alpha\,\omega(z) + \omega(z)\int dz'\,\frac{B(z')}{\omega(z')}\,,
\eeq
where $\alpha\in\mathbb{R}$ is an integration constant. The poles appearing inside the integral in eq.~\eqref{eq:solution} are determined by the poles 
in the coefficient $A(z)$. Thus in our case the solution for the master integrals can be expressed through (iterated) integrals with poles at most
 at $z=0$ and $z=\pm1$. Integrals of this form lead naturally to harmonic polylogarithms~\cite{Remiddi:1999ew}, defined recursively through the iterated integrals
\beq\label{eq:hpldef}
H(a_1,\ldots,a_n;z) = \int_0^zdt\,f(a_1,t)\,H(a_2,\ldots,a_n;t)\,,
\eeq
where $a_i\in\{-1,0,1\}$ and
\beq
f(-1,t) = \frac{1}{1+t}\,,\qquad f(0,t) = \frac{1}{t}\,,\qquad f(1,t) = \frac{1}{1-t}\,.
\eeq
 In the case where all the $a_i$ are zero, the integral~\eqref{eq:hpldef} is divergent, and we define instead
\beq
H(\vec0_n;z) = \frac{1}{n!}\log^nz\,.
\eeq
In this way we can express all the master integrals, up to the desired order in $\eps$, in terms of harmonic polylogarithms up to weight four, which can be evaluated numerically in a fast
 and accurate way~\cite{  Gehrmann:2001pz,Vollinga:2004sn,Maitre:2005uu,Maitre:2007kp,Buehler:2011ev}.

In order to fully determine the solutions of the differential equations, we need to fix the integration constants. This can be achieved by requiring the
master integrals to take particular values, computed separately by other means, for some special value of $z$. Since $0<z<1$, there are two natural choices 
for such special values. The case $z=0$ physically corresponds to the situation where the produced particle $X$ is massless. However, the limit $z\to0$ is in general
 not smooth, as the amplitude might develop new infrared poles in this limit. For this reason we choose the point $z=1$ as an initial condition to the differential
 equation, which corresponds to the \emph{soft limit} in which the momenta of all the final-state partons vanish. As this limit is important not only as an initial condition 
for the differential equations, we review the structure of the phase space integrals in the soft limit in the next section.

\subsection{Soft limits of master integrals}
\label{sec:softie}
In the previous section we argued that the point $z=1$ is a good initial condition for the differential equations satisfied by the master integrals. 
There is however another reason why the point $z=1$ deserves special attention. Indeed,
some of the master integrals contain simple poles at $z=1$, and without loss of generality we can always write such integrals in the form
\beq
\textbf{F}(z,\eps)=\sum_{n\in  \mathbb{Z}^\times} \frac{F_n(z,\eps)}{(1-z)} {}_{1+n \eps},
\eeq
where $n$ is an integer and the function $F_n(z,\eps)$ is finite at $z=1$. 
We can isolate the divergence associated to the soft singularity by expanding
\beq\label{eq:plus_exp}
(1-z)^{-1+n \eps}=\frac{\delta(1-z)}{n\eps} + \sum_{k=0}^\infty \frac{(n\eps)^k}{k!} \Di{k}(1-z) 
\eeq
where 
\beq
\Di{k}(x)= \left[\frac{\log^k(x)}{x}\right]_+ 
\eeq
and the $+$ indicates the common plus-prescription,
\beq
\int_0^1 dx\,\Di{k}(x)\,f(x) = \int_0^1 dx\,\frac{\log^k(x)}{x}\,\Big[f(x)-f(0)\Big]\,.
\eeq
It then becomes apparent that, if we want to know $\textbf{F}(z,\eps)$ up to a certain order in the $\eps$ expansion, we need to know $F_n(1,\epsilon)$ 
at one order higher in the expansion in $\eps$ than $F_n(z,\epsilon)$, since $F_n(1,\eps)$ is multiplied by the pole in eq.~\eqref{eq:plus_exp}. 
More generally, even if a given master integral has no pole as $z$ approaches $1$, when appearing inside an amplitude it might get
 multiplied by a coefficient such that the product develops at most a simple pole at $z=1$. Hence, we have to use eq.~\eqref{eq:plus_exp} to perform the $\eps$ expansion and therefore, by the same logic, we always need to know the values of the master integrals at $z=1$ to one order higher in the $\eps$ expansion. Thus, there is
 an independent physics motivation to study the soft limits of the master integrals in some detail.

We define the soft limit of a master integral $\textbf{F}(z,\eps)$ as the (unique) function
\beq
\textbf{F}^S(z,\eps) = \sum_{n\in\mathbb{Z}^\times}\frac{F_n(\eps)}{(1-z)}{}_{\alpha+n\eps}\,,\qquad \alpha\in\mathbb{Z}_{\ge1}\,,
\eeq
such that
\beq\label{eq:init_cond}
\lim_{z\to1}\frac{\textbf{F}(z,\eps)}{\textbf{F}^S(z,\eps)} = 1\,.
\eeq
We stress that the right-hand side of eq.~\eqref{eq:init_cond} is valid to all orders in $\eps$. 
In addition we define the hard part $\textbf{F}^H(z,\eps)$ of a master integral by
\beq
\textbf{F}(z,\eps) = \textbf{F}^S(z,\eps) + \textbf{F}^H(z,\eps)\,.
\eeq
In sections~\ref{sec:RR} and~\ref{sec:RV} we will determine the soft limits of the master integrals directly to all orders in $\eps$, without the help of
 the differential equations. Our knowledge of the soft limits combined with eq.~\eqref{eq:init_cond} as an initial condition for the 
differential equations then 
allow us, at least in principle, to compute the analytic expressions for the master integrals up to any order in $\eps$. In the rest of this paper we perform this task explicitly
 for all master integrals up to transcendental weight 4.


\section{Double-real master integrals}
\label{sec:RR}

\subsection{Definitions and conventions}
In this section we compute the analytic expressions for the master integrals for the double-real emission contributions. 
In ref.~\cite{Anastasiou:2002yz} it was shown that all phase space integrals for the production of a massive particle and two massless 
particles can be reduced to a linear combination of 18 master integrals, which we choose as follows:

%

\begin{equation}
 \Xone{27} = \int  d\Phi_3 = s^{1-2\eps}\,\P(\eps)\,\XOne(z,\eps)\,,
 \end{equation}

\begin{equation}
\Xthree{27} = \int \frac{ d\Phi_3}{ s_{234}  s_{134}} =s^{-1-2\eps}\,\P(\eps)\,\XThree(z,\eps)\,,
\end{equation}

\beq
\Xfour{27} =  \int \frac{ d\Phi_3}{ s_{14}  s_{23} } =s^{-1-2\eps}\,\P(\eps)\,\XFour(z,\eps)\,,
\eeq

\beq
\Xfive{27} =  \int \frac{ d\Phi_3}{ s_{234}  s_{23} s_{13} s_{134}} =s^{-3-2\eps}\,\P(\eps)\,\XFive(z,\eps)\,,
\eeq

\beq
\Xsix{27} =  \int \frac{ d\Phi_3}{ s_{234}  s_{24} s_{13} s_{134}} =s^{-3-2\eps}\,\P(\eps)\,\XSix(z,\eps)\,,
\eeq

\beq
\Xeight{27} =  \int \frac{ d\Phi_3}{ s_{13}  s_{23} s_{14} s_{24}} =s^{-3-2\eps}\,\P(\eps)\,\XEight(z,\eps)\,,
\eeq

\beq
 \Xten{27} =  \int \frac{ d\Phi_3}{ s_{123}  s_{124} } =s^{-1-2\eps}\,\P(\eps)\,\XTen(z,\eps)\,,
\eeq

\beq
 \Xeleven{27} =  \int \frac{ d\Phi_3}{ s_{123}  s_{124} s_{14}s_{13}} =s^{-3-2\eps}\,\P(\eps)\,\XEleven(z,\eps)\,,
\eeq

\beq
 \Xtwelve{27} =  \int \frac{ d\Phi_3}{ s_{14}  s_{23} } =s^{-1-2\eps}\,\P(\eps)\,\XTwelve(z,\eps)\,,
\eeq

\beq
\Xthirteen{27} =  \int \frac{ d\Phi_3}{ s_{123}  s_{124} s_{14}s_{23}} =s^{-3-2\eps}\,\P(\eps)\,\XThirteen(z,\eps)\,,
\eeq


\beq
\Xfourteen{27} = \int d\Phi_3 s_{34} = s^{2-2\eps}\,\P(\eps)\,\XNineNine(z,\eps)\,,
\eeq

\beq
\Xfifteen{27} = \int \frac{ d\Phi_3 s_{23}}{ s_{234} s_{123}} =s^{-2\eps}\,\P(\eps)\,\XFifteen(z,\eps)\,,
\eeq

\beq
\Xsixteen{27} = \int \frac{ d\Phi_3 }{ s_{234} s_{123}} =s^{-1-2\eps}\,\P(\eps)\,\XSixteen(z,\eps)\,,
\eeq

\beq
\Xseventeen{27} = \int \frac{ d\Phi_3 }{ s_{13} s_{124}} =s^{-1-2\eps}\,\P(\eps)\,\XSeventeen(z,\eps)\,,
\eeq

\beq
\Xeighteen{27} = \int \frac{ d\Phi_3 }{ s_{13} s_{124} s_{134}} =s^{-2-2\eps}\,\P(\eps)\,\XEighteen(z,\eps)\,,
\eeq

\beq
\Xtwenty{27} = \int \frac{ d\Phi_3 }{ s_{14} s_{23} s_{234}s_{124}} =s^{-3-2\eps}\,\P(\eps)\,\XTwenty(z,\eps)\,,
\eeq

\beq
\Xtwentyfive{27} =\int \frac{ d\Phi_3 }{ s_{34} s_{13} s_{234}s_{123}} =s^{-3-2\eps}\,\P(\eps)\,\XTwentyFive(z,\eps)\,,
\eeq

\beq
\Xthirty{27} = \int \frac{ d\Phi_3 }{ s_{34} s_{13} s_{24}} =s^{-2-2\eps}\,\P(\eps)\,\XThirty(z,\eps)\,,
\eeq
\vskip1cm
The solid lines denote the massive particle and dotted lines represent numerator factors. 
In all integrals we have pulled out an overall dimensionful  scaling factor as well as an overall normalization
\beq
\mathcal{P}(\eps)= \frac{1}{2}\, \frac{1}{(4\pi)}{}_{3-2\eps}\, \frac{\Gamma(1-\eps)^2}{\Gamma(2-2\eps)^2}\,.
\eeq
Note that
the integral $\XOne$ is just the phase space volume, which can be written to all orders in $\eps$ as
\beq
\label{eq:psvolume}
\XOne(z,\eps)  = (1-z)^{3-4\epsilon}  \frac{\Gamma(2-2\eps)^2}{\Gamma(4-4\eps)}  {}_2F_1(1-\epsilon,2-2\epsilon;4-4\epsilon;1-z) \, .
\eeq
For the other integrals no similar all-orders formulae are known to us. We therefore compute these integrals using the differential equation technique. It turns out that the system of differential equations for the double-real master integrals are triangular order by order in $\eps$, except for $\XOne$ and $\XNineNine$, which can be decoupled order by order by replacing $\XNineNine$ by the linear combination
\beq
\XFourteen = \frac{1}{2}\big[(1+z)\,\XOne - \XNineNine\big]\,.
\eeq
After this change of variables the system is triangular order by order in $\eps$, and we can immediately solve for the master integrals, provided that
we know the soft limits of the master integrals which serve as an initial condition. The computation of the soft limits will be the topic of the rest of this section.

\subsection{Soft limits of double-real master integrals}
\label{sec:soft_RR}
To evaluate the soft limits of the double-real master integrals we use the following exact parametrization for the $2\to 3$ phase space,
\begin{eqnarray}
\int d\Phi_3 & = & \frac{(2\pi)^{-3+2\epsilon}}{16\Gamma(1-2\epsilon)} \int_0^1 dx_2 \(\prod_{i=1}^4 dy_i\) \delta\(\sum_{i=1}^4 y_i-1\)    \\
&& \quad  \quad \quad \quad \quad \quad \times \quad \quad \left(\frac{s\bar z^3\kappa^{4} }{2-\kappa}\right)  \left(s^2 \bar z^4 \kappa^4 \sin^2(\pi x_2) \prod_{i=1}^4 y_i   \right)^{-\epsilon}\,,    \nonumber 
\end{eqnarray}
with $\bar{z}=1-z$. In this parametrization the propagators of the massless  partons read
\begin{eqnarray}
&&s_{13}  =-s\bar z\kappa y_1,  \qquad  s_{23}  =-s\bar z\kappa y_2, \nn\\ 
&& s_{14}  =-s\bar z\kappa y_3,  \qquad s_{24}  =-s\bar z\kappa y_4, \label{eq:s_to_y}\\
&& \qquad \qquad \quad s_{34}   = s\bar z\kappa^2 \xi. \nn
\end{eqnarray}
where 
\beq
\xi=\bar z \(y_1y_4+y_2y_3+2\cos(x_2\pi)\sqrt{y_1y_4y_2y}\)\,,
\eeq
and
\begin{equation}
\kappa=\frac{1-\sqrt{1-4\xi}}{2\xi}\quad \in [1,2).  
\end{equation}
This parametrization may be derived from the ``energies and angles parametrization'' of ref.~\cite{Anastasiou:2010pw}
via the following transformation
\beq\label{eq:y_param_1}
y_1  =  x_1 x_3,   \qquad y_2  =  x_1 \bar x_3, \qquad y_3  =  \bar x_1 x_4, \qquad y_4  =  \bar x_1\bar x_4\,,
\eeq
with $\bar x_i = 1-x_i$.
In addition, the master integrals depend on the following denominators
\beq\bsp
s_{134}  =  s_{13}+s_{14}+s_{34}\,,&\qquad s_{234}  =  s_{23}+s_{24}+s_{34}\,, \\
s_{123}  =  s_{12}+s_{23}+s_{13}\,, &\qquad s_{124}  =  s_{12}+s_{24}+s_{14}\,.
\esp\eeq

We now take the soft limit, $z\to1$. It is easy to see that in the soft limit we have the relations
\beq\bsp
\lim_{z\to 1} s_{123}  = \lim_{z\to 1} s_{124} = s_{12}\,,&\qquad\lim_{z\to 1} \kappa =1\,,\\
\lim_{z\to 1} s_{134}  =  s_{13}+s_{14}\,,\qquad\, &\qquad\lim_{z\to 1} s_{234}  =  s_{23}+s_{24}\,.
\esp\eeq
If we now perform the change of variables
\beq\label{eq:y_param_2}
y_1  =  t_1 t_3,    \qquad y_2  =  \bar t_1 t_4, \qquad y_3  =  t_1 \bar t_3, \qquad y_4  =  \bar t_1\bar t_4\,.
\eeq
It is then easy to see from eqs.~\eqref{eq:s_to_y} and~\eqref{eq:y_param_2} that in the soft limit
all the invariants, except for $s_{34}$, take a fully factorized form. In particular
the terms $s_{13}+s_{14}$ and $s_{23}+s_{24}$ effectively reduce to $t_1$ and $\bar t_1$ respectively. 
This construction therefore allows one to derive all the soft limits of almost all master integrals,
with the exception of $\XTwentyFive$ and $\XThirty$, in terms of simple Beta-functions,
\beq
B(x,y) = \int_0^1dt\,t^{x-1}\,(1-t)^{y-1} = \frac{\Gamma(x)\,\Gamma(y)}{\Gamma(x+y)}\,.
\eeq 
The integral $\XTwentyFive$ may be trivialized in a similar way in the ``hierarchical parametrization'' of ref.~\cite{Anastasiou:2010pw}. 
Unfortunately, we are not aware of any parametrization which allows one to factorize all denominators of $\XThirty$ simultaneously. 
We proceed by expressing the integral in terms of energies and angles, i.e., by writing $p_i=E_i(1,\vec{n}_i)$, where 
$\vec{n}_i$ is a unit vector in the direction of the spatial component  of $p_i$. 
The integral over the energies of $p_3$ and $p_4$ yields just another Beta-function,
and we arrive at
\beq
\P(\eps)\XThirty^S =4 (4\pi)^{-5+4\eps} \frac{B(-2\eps,-2\eps)} {(1-z)^{1+4\eps}}  \int \frac{d\Omega_3^{(d-1)}d\Omega_4^{(d-1)}}{(1+\cos\theta_{13})(1-\cos\theta_{14})(1-\cos\theta_{34})} \, ,
\eeq
where $d\Omega_i^{(d-1)}$ is differential volume element of the $d-1$ dimensional solid angle of $p_i$ and
the angles are defined as $\cos\theta_{ij}=\vec{n}_i.\vec{n}_j$.
We first use the result~\cite{vanNeerven:1985xr}
\bea
\Omega_{11}&=&\int \frac{d\Omega_4^{(d-1)}}{(1-\cos\theta_{14})(1-\cos\theta_{34})} \nn\\
 &=& \Omega^{(d-3)}\int_0^\pi \frac{d\theta_{13} (\sin\theta_{13})^{d-3}  d\phi_{34} (\sin\phi_{34})^{d-4}  }{ (1-\cos\theta_{14}) (1-\cos\theta_{14}\cos\theta_{13}-\cos\phi_{34}\sin\theta_{14}\sin\theta_{13} )}\\
 &=& -\frac{1}{4} (4\pi)^{1-\eps} \frac{\Gamma(-\eps)}{\eps\Gamma(-2\eps)}\, {}_2F_1\(1,1;1-\eps;\frac{1+\cos\theta_{13}}{2}\).\nn
\eea
The integral then becomes 
\beq\bsp\label{eq:X30S}
\P(\eps)\XThirty^S= -\frac{(4\pi)^{-4+3\eps}}{ (1-z)^{1+4\eps}} &\frac{\Gamma(-\eps)\Gamma(-2\eps)}{\eps\Gamma(-4\eps)} \Omega^{(d-2)}\\
&\times \int_0^\pi \frac{d\theta_{13}(\sin\theta_{13})^{d-3}}{1+\cos\theta_{13}}  {}_2F_1\(1,1;1-\eps;\frac{1+\cos\theta_{13}}{2}\).
\esp\eeq
Changing variables to $y=\frac{1+\cos\theta_{13}}{2}$, eq.~\eqref{eq:X30S} may be written as
\bea
\P(\eps)\XThirty^S &=& \frac{(4\pi)^{-3+2\eps}}{ (1-z)^{1+4\eps}} \frac{\Gamma(-2\eps)}{\eps^2\Gamma(-4\eps)}  \int_0^1 dyy^{-1-\eps}(1-y)^{-\eps} {}_2F_1\(1,1;1-\eps;y\) \nn\\
                &=& \frac{1}{2}\frac{(4\pi)^{-3+2\eps}}{ (1-z)^{1+4\eps}} \frac{\Gamma(-\eps)^2}{\eps^2\Gamma(-4\eps)}\, {}_3F_2\(1,1,-\eps;1-\eps,1-2\eps;1\),
\eea
where we used the recursive definition of the ${}_pF_q$ function.
Hence we arrive at expressions valid to all orders in $\eps$ for the soft limits of all the double-real master integrals. We observe that in all
 cases, except for $\XThirty^S$, the results are proportional to the soft limit of the phase space volume, 
\beq
\XOne^S(z,\eps)  =  (1-z)^{3-4\epsilon}  \frac{\Gamma(2-2\eps)^2}{\Gamma(4-4\eps)}\,,
\eeq
the constant of proportionality being a rational function of $z$ and $\eps$. We therefore define
\beq
\X_i^S(z,\eps) = \S_i(z,\eps)\,\XOne^S(z,\eps)\,.
\eeq
In this normalization the results for the soft limits of the master integrals read
\beq
\SOne(z,\eps) = \STen(z,\eps) = \SFourteen(z,\eps) = 1\,,
\eeq
\beq
\SThree(z,\eps)  = {\frac {2(3-4\eps)}{ \left( 1-2\eps \right) {(1-z)}^{
2}}}\,,
\eeq
\beq
\SFour(z,\eps)  =\SEleven(z,\eps)=\STwelve(z,\eps)=\SThirteen(z,\eps)=2{\frac {(1-2\eps)(3-4\eps) }{{\eps}^{2}
{(1-z)}^{2}}}\,,
\eeq
\beq
\SFive(z,\eps)  =\SSix(z,\eps)    =-2\,{\frac { \left(1-2\eps \right)  \left(3-4\eps
 \right)  \left(1-4\eps \right) }{{\eps}^{3}{(1-z)}^{4}
}}\,,
\eeq
\beq
\SEight(z,\eps)  =-8\,{\frac { \left(1-2\eps \right)  \left(3-4\eps
 \right)  \left(1-4\eps \right) }{{\eps}^{3}{(1-z)}^{4}
}}\,,
\eeq
\beq
\SFifteen(z,\eps)  =\frac{1}{2}\,,
\eeq
\beq
\SSixteen(z,\eps)  =-{\frac {3-4\eps}{(1-2\eps)\,(1-z)}}\,,
\eeq
\beq
\SSeventeen(z,\eps)  =
{\frac {3-4\eps}{\eps\,(1-z)}}\,,
\eeq
\beq
\SEighteen(z,\eps)  ={\frac {(1-2\eps)(3-4\eps) }{{\eps}^{2}
{(1-z)}^{2}}}\,,
\eeq
\beq
\STwenty(z,\eps)  ={\frac { \left( 1-2\eps \right)  \left(3-4\eps \right) 
 \left(1-4\eps \right) }{{\eps}^{3}{(1-z)}^{3}}}\,,
\eeq
\beq
\STwentyFive(z,\eps)  =-2\,{\frac { \left(1-2\eps \right)  \left(3-4\eps
 \right)  \left(1-4\eps \right) }{{\eps}^{3}{(1-z)}^{4}
}}\,,
\eeq
\beq\bsp\label{eq:S30}
\SThirty(z,\eps)  = &\,-4\frac{(1-2\eps)(3-4\eps)(1-4\eps)}{\eps^3(1-z)^4}{}_3 F_2 \(1,1,-\eps;1-\eps,1-2\eps;1\)\\
=&\, \frac{1}{(1-z)^4}\Bigg[-\frac{18}{\epsilon^3}+\frac{132}{\epsilon^2}+\frac{1}{\epsilon}\left(12\zeta_2-288\right)-88\zeta_2+60\zeta_3+192\\
&\,\qquad\qquad\,\,+\left(\frac{372}{5}\zeta_2^2+192\zeta_2-440\zeta_3\right)\epsilon\\
&\,\qquad\qquad\,\,+\left(-\frac{2728}{5}\zeta_2^2+24\zeta_3\zeta_2-128\zeta_2+444\zeta_5+960\zeta_3\right)\epsilon^2
\\
&\,\qquad\qquad\,\,+\ord(\eps^3)\Bigg]\,.
\esp\eeq

One might wonder whether our result for $\SThirty$, eq.~\eqref{eq:S30}, can be expressed to all orders in $\eps$ through ratios of $\Gamma$ functions only.
 Without loss of generality we can assume that in that case the ${}_3F_2$ function entering $\SThirty$ could be written in the form,
\beq\label{eq:ansatz_S30}
R(\eps)\,\prod_k \Gamma(1\pm n_k\eps)^{m_k}\,,\qquad m_k,n_k\in\mathbb{Z}^\times\,,
\eeq
where $R(\eps)$ is a rational function. In the following we argue that the form~\eqref{eq:ansatz_S30} can be excluded for $\SThirty$, i.e., that 
${}_3 F_2 \(1,1,-\eps;1-\eps,1-2\eps;1\)$ cannot be reduced to a ratio of $\Gamma$ functions only.
Indeed, if we assume that the  ${}_3 F_2$ function in eq.~\eqref{eq:S30} can be written in the form~\eqref{eq:ansatz_S30}, then, because of the well-known relation
\beq
\Gamma(1+n\eps) = \exp\left[-\gamma_E\eps+\sum_{k=2}^\infty\frac{(-n\eps)^k}{k}\,\zeta_k\right]\,,
\eeq
its $\eps$ expansion will to all orders in $\eps$ only involve zeta values of depth one, i.e., it will be possible to write its $\eps$ expansion without
 \emph{multiple} zeta values (MZVs). Multiple zeta values are known to satisfy many identities among themselves, and it is conjectured that \emph{all} 
the identities among MZVs are generated by the double shuffle relations. Using the double-shuffle relations, it can be shown that the first time a MZV
 cannot be written as a polynomial of MZVs of depth one is at weight 8, and this irreducible MZV can be chosen to be $\zeta_{5,3}$. Using {\tt HypExp}
 we can perform the expansion of ${}_3 F_2 \(1,1,-\eps;1-\eps,1-2\eps;1\)$ up to $\ord(\eps^8)$. The result obtained by {\tt HypExp} is a complicated 
combination of MZVs of weight 8. Using the PSLQ algorithm we fit this combination of MZVs to a basis of MZVs of weight 8, which we choose to be 
$\{\zeta_2^4, \zeta_2\,\zeta_3^2,\zeta_3\,\zeta_5, \zeta_{5,3}\}$. We obtain
\beq
{}_3 F_2 \(1,1,-\eps;1-\eps,1-2\eps;1\)_{|\eps^8} = -\frac{7489}{125}\zeta_2^4-2\zeta_3^2\,\zeta_2-74\zeta_3\,\zeta_5-\frac{44}{5}\zeta_{5,3}\,.
\eeq
We see that at $\ord(\eps^8)$ the expansion of ${}_3 F_2 \(1,1,-\eps;1-\eps,1-2\eps;1\)$ involves an irreducible MZV of weight 8.
 Conversely, we checked that if we remove this MZV from our basis at weight 8, the PSLQ algorithm fails to converge. We thus conclude
 that the ${}_3 F_2$ function appearing in $\SThirty$ cannot be written as a ratio of $\Gamma$ functions of the form~\eqref{eq:ansatz_S30}.

Having at our disposal both the differential equations and the soft limits for all the master integrals, we can in principle solve the differential equations
 to any order in $\eps$. We have explicitly performed this task and computed the $\eps$ expansion of all the master integrals up to transcendental weight 4. 
As the results are rather lengthy, we do not show them here explicitly but collect them in appendix~\ref{sec:results_RR}. The results only involve harmonic
 polylogarithms up to weight 4. In ref.~\cite{Duhr:2011zq} it was shown that up to weight 4 almost all harmonic polylogarithms can be expressed through classical
 polylogarithms only, and that only three new functions are needed. In our case only one new function appears, which we choose to correspond to the harmonic polylogarithm
\beq
\textrm{Li}_{2,2}(-1,z) = -H(0,1,0,-1;z)\,.
\eeq
We have checked that our results agree with the results of ref.~\cite{Anastasiou:2002yz} up to terms in the $\eps$ expansion 
with transcendental weight at most three. In addition we have checked all our results numerically for a few nontrivial values of $z$.
For this purpose we used the method of non-linear mappings introduced in ref.~\cite{Anastasiou:2010pw}.

\section{Real-virtual master integrals}
\label{sec:RV}

\subsection{Master integral definitions}

In ref.~\cite{Anastasiou:2002yz} six master integrals for the real-virtual contributions to a $2\to1$
partonic cross section were identified. Defining the prefactor 
\beq
\Q(\eps)=\frac{1}{2} \frac{(4\pi)^\eps}{4\pi} \frac{\Gamma(1+\eps)\Gamma(1-\eps)}{\Gamma(2-2\eps)},
\eeq
as well as the one-loop integrals
\bea
\textbf{Bub}(p^2)&=&\int \frac{d^Dk}{i\pi^{D/2}} \frac{1}{k^2(k+p)^2}, \nn \\
\textbf{Tri}(s_{23},s_{123})&=&\int \frac{d^Dk}{i\pi^{D/2}} \frac{1}{k^2(k+p_1)^2(k+p_{123})^2}, \\
\textbf{Box}(s_{12},s_{23},s_{123})&=&\int \frac{d^Dk}{i\pi^{D/2}} \frac{1}{k^2(k+p_1)^2(k+p_{12})^2(k+p_{123})^2}\nn ,
\eea
the real-virtual master integrals may be defined as the following two-body phase space integrals,
\beq
\YA{27} = \int  d\Phi_2 \textbf{Bub}(s_{13}) = s^{-2\eps}\Q(\eps)\textbf{Y}_1(z,\eps), 
\eeq
\beq
\YB{27} = \int  d\Phi_2 \textbf{Bub}(m_X^2) =s^{-2\eps}\Q(\eps) \textbf{Y}_2(z,\eps), 
\eeq
\beq
\YE{27} = \int  d\Phi_2 \textbf{Bub}(s_{12}) =s^{-2\eps}\Q(\eps) \textbf{Y}_5(z,\eps), 
\eeq
\beq
\YC{27} = \int  d\Phi_2 \textbf{Tri}(s_{13},m_X^2) =s^{-1-2\eps}\Q(\eps) \textbf{Y}_3(z,\eps), 
\eeq
\beq
\YD{27} = \int  d\Phi_2 \frac{\textbf{Box}(s_{13},s_{12},m_X^2)}{s_{23}} =s^{-3-2\eps}\Q(\eps) \textbf{Y}_4(z,\eps), 
\eeq
\beq
\YF{27} = \int  d\Phi_2 \textbf{Box}(s_{13},s_{23},m_X^2) =s^{-2-2\eps}\Q(\eps) \textbf{Y}_6(z,\eps) .
\eeq
\vskip1cm

\subsection{Soft limits of real-virtual master integrals}
\label{sec:soft_RV}
For the evaluation of the soft limits of the real-virtual master integrals we employ the 
strategy presented in ref.~\cite{Buehler:2012cu}. This strategy involves the following steps:
\begin{enumerate}
 \item Substitute known expressions for the one-loop master integrals which are valid to all orders in $\eps$.
 \item Apply analytic continuation formulae to the hypergeometric functions which express the box integrals, if required.
 \item Evaluate the integrals at $z=1$.
\end{enumerate}
We find that the soft limits of all master integrals can then be trivially evaluated as Beta-functions:

\beq
\Y_1^S(z,\eps) = (1-z)^{1-3\eps}\frac{\Gamma(1-\eps)\Gamma(1-2\eps)}{\eps\,\Gamma(2-3\eps)}\,,
\eeq

\beq
\Y_2^S(z,\eps) =(1-z)^{1-2\eps}\,\cos(\pi\eps)\,\frac{\Gamma(1-\eps)^2}{\eps\,\Gamma(2-2\eps)}\,,
\eeq

\beq
\Y_5^S(z,\eps) =\Y_2^S(z,\eps)\,,
\eeq

\beq
\Y_3^S(z,\eps) =0\,,\\
\eeq

\beq
\Y_4^S(z,\eps) =-3(1-z)^{-1-3\epsilon}\frac{\Gamma(2-2\epsilon)\Gamma(1-\epsilon)}{\epsilon^3\,\Gamma(1-3\epsilon)}\,,
\eeq

\beq
\Y_6^S(z,\eps) =-2(1-z)^{-1-4\eps}\,\cos(\pi\epsilon)\,\frac{\Gamma(1-2\epsilon)\Gamma(2-2\epsilon)\Gamma(1-\epsilon)\Gamma(1+\epsilon)}{\epsilon^3\,\Gamma(1-4\epsilon)}\,.
\eeq
Note that only the real part of the master integrals is shown, as the imaginary part never enters the computation of a physical observable.

As in the double-real case, we list the results for generic $z$ that we obtained by solving the differential equations in appendix \ref{sec:RVappendix}. The expressions
agree with the results of ref.~\cite{Anastasiou:2002yz} up to terms in the $\eps$ expansion with transcendental weight at most three. The new orders were checked numerically using
the approach of \cite{Buehler:2012cu}. This strategy is as follows:
\begin{enumerate}
 \item Substitute known expressions for the one-loop master integrals which are valid to all-orders in $\eps$.
 \item Apply analytic continuation formulae to the hypergeometric functions which express the box integrals, if required.
 \item Expand the hypergeometric in terms of polylogarithmic functions.
 \item Expand the real emission singularities (these are factorized here) in terms of delta- and plus-distributions.
 \item Evaluate the coefficients numerically.  
\end{enumerate}

\section{Conclusions}
\label{sec:Conclusions}

We have given results for all double-real and real-virtual master integrals needed in the calculation of a generic $2\to 1$ inclusive scattering
cross section. 
The analytic results for the master integrals are provided in Maple input form as ancillary material to the arXiv submission.
The previously known expressions have been supplemented with one more order in the $\eps$ expansion. Together with the previously
published double-virtual results~\cite{Gehrmann:2005pd} listed in appendix \ref{sec:Results_VV}, all contributions to a \n3lo cross section involving
NNLO master integrals can thus be computed, as outlined in section \ref{sec:NNNLO}.

We have divided the master integrals into a soft and a hard part, each of which is given separately. This separation is motivated in
a twofold way. First, the soft limits provide a natural boundary condition to solve the differential equations for the master integrals. Second, 
divergences due to soft radiation have to be subtracted, leaving poles at $z=1$ and thus requiring the knowledge of the master integral to one order higher
in the $\eps$ expansion, as explained in section \ref{sec:softie}.

We have found the method of differential equations to be very adequate for this type of task, and see no obvious problems in applying
the same techniques to integrals of higher complexity.
Such integrals could include the triple-real master integrals which constitute another part of a $2\to 1$ \n3lo cross section. In that case, the main difficulty might
be the vastly increased amount of master integrals, as well as a much more intertwined set of differential equations due to more complicated IBP identities for the 
triple-real topologies.
Once these difficulties overcome, one should be able to perform analytically all the phase space integrals appearing in $2\to 1$ cross sections at \n3lo, and thus, 
by combining them with the recently computed three-loop QCD form factor~\cite{Gehrmann:2010ue,Lee:2010cga}, arrive at the first prediction for an LHC observable computed
 perturbatively up to \n3lo.

\section*{Acknowledgements}
This work was supported by the ERC grant ``IterQCD'' and the Swiss National Foundation under contract SNF 200020-126632. All Feynman diagrams in
 this paper were drawn with the help of the axodraw package~\cite{axodraw}.

\appendix
\section{Results for the master integrals}
\label{sec:results}
In this appendix we collect the analytic results for the hard part of the real-virtual and double-real master integrals omitted throughout the main text. As the imaginary part of the master integrals does not enter the computation of any physical observable, we only show the real part.
For completeness we also include the analytic expressions for the two-loop virtual master integrals, which can easily be obtained from the 
expressions for the two-loop  form factor master integrals of ref.~~\cite{Gonsalves:1983nq,Kramer:1986sr,Gehrmann:2005pd}.

\subsection{Results for the double-virtual master integrals}
\label{sec:Results_VV}

\begin{align}
\cglass{27} &=  \delta(1-z)\,\frac{\Gamma(1+\eps)^2\,\Gamma(1-\eps)^4}{\eps^2\,\Gamma(2-2\eps)^2}\,,
\end{align}

\beq
\glass{27} = \delta(1-z)\,\cos(2\pi\eps)\, \frac{\Gamma(1+\eps)^2\,\Gamma(1-\eps)^4}{\eps^2\,\Gamma(2-2\eps)^2}\,.
\eeq

\begin{align}
 \suns{27} &= 
-\delta(1-z)\,\cos(2\pi\eps)\,\frac{\Gamma(1-\epsilon)^3\Gamma(1+2\epsilon)}{2(1-2\epsilon)\,\epsilon\,\Gamma(3-3\epsilon)}\,,
\end{align}

\begin{align}
 \tri{27} &=  \delta(1-z)\,\cos(2\pi\eps)\, \frac{\Gamma(1-2\epsilon)\Gamma(1-\epsilon)^2\Gamma(1+\epsilon)\Gamma(1+2\epsilon)}{2(1-2\epsilon)\epsilon^2\Gamma(2-3\epsilon)}\,,
\end{align}

\begin{align}
 \xtri{27} &= \delta(1-z)\,\cos(2\pi\eps)\,\Gamma(1+\eps)^2\,\Bigg\{
\frac{1}{\epsilon^4}-\frac{7}{\epsilon^2}\zeta_2-\frac{27}{\epsilon}\zeta_3
-\frac{57}{5}\zeta_2^2\\
\nonumber&+\Big[102\zeta_2\zeta_3-117\zeta_5\Big]\epsilon+\ord(\eps^2)\Bigg\}\,,
\end{align}

\subsection{Results for the real-virtual master integrals}
\label{sec:RVappendix}

\begin{align}
\Y_1^H(z,\eps) &=0\,,\\
\Y_2^H(z,\eps) &=(z^{-\eps}-1)\,\Y_2^S(z,\eps)=(z^{-\eps}-1)(1-z)^{1-2\eps}\,\cos(\pi\eps)\,\frac{\Gamma(1-\eps)^2}{\eps\,\Gamma(2-2\eps)}\,,\\
\Y_5^H(z,\eps) &=0\,,
\end{align}

\begin{align}
\Y_3^H(z,\eps) &=z^{-2 \epsilon }\Bigg\{\frac{1}{\epsilon }\Big[\text{Li}_2(z)+\frac{\log^2z}{2}-\zeta_2\Big]+-5\text{Li}_3(1-z)-4\text{Li}_3(z)\\
\nonumber&+4\text{Li}_2(z)\log z-5\text{Li}_2(z)\log(1-z)+2\zeta_2\log z+5\zeta_2\log(1-z)+\frac{\log^3z}{2}\\
\nonumber&-\frac{5}{2}\log^2(1-z)\log z+4\zeta_3-2\text{Li}_2(z)-\log^2z+2\zeta_2+\epsilon  \Big[2\zeta_2\text{Li}_2(z)\\
\nonumber&+14\text{Li}_4\left(1-\frac{1}{z}\right)-5\text{Li}_4(1-z)-4\text{Li}_4(z)+5\text{Li}_2(z)\log^2z+\frac{19}{2}\text{Li}_2(z)\log^2(1-z)\\
\nonumber&-14\text{Li}_2(z)\log(1-z)\log z-14\text{Li}_3(1-z)\log z-10\text{Li}_3(z)\log z\\
\nonumber&+19\text{Li}_3(1-z)\log(1-z)+14\text{Li}_3(z)\log(1-z)+7\zeta_2\log^2z-\frac{19}{2}\zeta_2\log^2(1-z)\\
\nonumber&+10\zeta_3\log z-14\zeta_3\log(1-z)+\frac{7\log^4z}{8}-\frac{7}{3}\log(1-z)\log^3z\\
\nonumber&+\frac{19}{3}\log^3(1-z)\log z-\frac{7}{2}\log^2(1-z)\log^2z-\frac{2}{5}\zeta_2^2+10\text{Li}_3(1-z)+8\text{Li}_3(z)\\
\nonumber&-8\text{Li}_2(z)\log z+10\text{Li}_2(z)\log(1-z)-4\zeta_2\log z-10\zeta_2\log(1-z)-\log^3z\\
\nonumber&+5\log^2(1-z)\log z-8\zeta_3+\ord(\eps^2)\Big]\Bigg\}\,,
\end{align}

\begin{align}
\Y_4^H(z,\eps) &=(1-z)^{-1-3\epsilon}\Bigg\{-\frac{1}{\epsilon ^2}3\log z+\frac{1}{\epsilon }\Big[-\text{Li}_2(z)+\log^2z+\zeta_2+6\log z\Big]\\
\nonumber&+\text{Li}_3(1-z)+2\text{Li}_3(z)+\text{Li}_2(z)\log(1-z)-\zeta_2\log(1-z)+8\zeta_2\log z\\
\nonumber&+\frac{1}{2}\log z\log^2(1-z)-2\zeta_3+2\text{Li}_2(z)-2\log^2z-2\zeta_2+\epsilon\Big[6\zeta_2\text{Li}_2(z)-2\text{Li}_2(z)^2\\
\nonumber&-10\text{Li}_4\left(1-\frac{1}{z}\right)-9\text{Li}_4(1-z)-6\text{Li}_4(z)-2\text{Li}_2(z)\log^2z-\frac{1}{2}\text{Li}_2(z)\log^2(1-z)\\
\nonumber&+8\text{Li}_3(z)\log z-\text{Li}_3(1-z)\log(1-z)-10\text{Li}_3(z)\log(1-z)-5\zeta_2\log^2z\\
\nonumber&+\frac{1}{2}\zeta_2\log^2(1-z)+10\zeta_2\log(1-z)\log z+8\zeta_3\log z+10\zeta_3\log(1-z)-\frac{7}{12}\log^4z\\
\nonumber&+\frac{5}{3}\log(1-z)\log^3z-\frac{1}{3}\log^3(1-z)\log z-\frac{5}{2}\log^2(1-z)\log^2z-\frac{8}{5}\zeta_2^2-2\text{Li}_3(1-z)\\
\nonumber&-4\text{Li}_3(z)-2\text{Li}_2(z)\log(1-z)+2\zeta_2\log(1-z)-16\zeta_2\log z-\log z\log^2(1-z)+4\zeta_3\Big]\\
\nonumber&+\ord(\eps^2)
\Bigg\}\,,
\end{align}

\begin{align}
\Y_6^H(z,\eps) &=(1-z)^{-1-4\eps}\Bigg\{-\frac{1}{\epsilon ^2}2\log z+\frac{1}{\epsilon }\Big[2\text{Li}_2(z)-2\zeta_2+4\log z\Big]+2\text{Li}_3(1-z)\\
\nonumber&+4\text{Li}_3(z)+2\text{Li}_2(z)\log(1-z)-2\zeta_2\log(1-z)+\frac{2}{3}\log^3z+\log^2(1-z)\log z\\
\nonumber&-4\zeta_3+4\zeta_2-4\text{Li}_2(z)+\epsilon\Big[12\zeta_2\text{Li}_2(z)-8\text{Li}_2(z)^2-20\text{Li}_4\left(1-\frac{1}{z}\right)\\
\nonumber&-22\text{Li}_4(1-z)-20\text{Li}_4(z)-8\text{Li}_2(z)\log^2z+\text{Li}_2(z)\log^2(1-z)+24\text{Li}_3(z)\log z\\
\nonumber&+2\text{Li}_3(1-z)\log(1-z)-20\text{Li}_3(z)\log(1-z)-2\zeta_2\log^2z-\zeta_2\log^2(1-z)\\
\nonumber&+20\zeta_2\log(1-z)\log z+8\zeta_3\log z+20\zeta_3\log(1-z)-\frac{4}{3}\log^4z\\
\nonumber&+\frac{10}{3}\log(1-z)\log^3z+\frac{2}{3}\log^3(1-z)\log z-5\log^2(1-z)\log^2z+4\zeta_2^2\\
\nonumber&-4\text{Li}_3(1-z)-8\text{Li}_3(z)-4\text{Li}_2(z)\log(1-z)+4\zeta_2\log(1-z)-\frac{4}{3}\log^3z\\
\nonumber&-2\log^2(1-z)\log z+8\zeta_3\Big]+\ord(\eps^2)
\Bigg\}\,,
\end{align}

\subsection{Results for the double-real master integrals}
\label{sec:results_RR}
\begin{align}
\XOne^H(z,\eps)& = z\log z-\frac{1}{6}(1-z)\left(z^2-5z-2\right)+\epsilon\Big[\frac{2}{3}(1-z)\left(z^2-5z-2\right) \log(1-z)\\
\nonumber& + 4z\text{Li}_2(z)-4z\zeta_2-\frac{1}{2}z\log^2z-\frac{1}{2}z(z+2)\log z-\frac{5}{36}(1-z)\left(4z^2-17z-5\right)\Big]\\
\nonumber&+\epsilon ^2\Big\{\Big[-8z\log z-\frac{4}{3}(1-z)\left(z^2-5z-2\right)\Big] \log^2(1-z)\\
\nonumber& + \Big[16z\zeta_2-16z\text{Li}_2(z)-\left(1-z^2\right)\log z+\frac{5}{9}(1-z)\left(4z^2-17z-5\right)\Big] \log(1-z)\\
\nonumber&-16z\text{Li}_3(1-z)-8z\text{Li}_3(z)+2z\text{Li}_2(z)\log z+8z\zeta_3+2z\zeta_2\log z+\frac{1}{6}z\log^3z\\
\nonumber&-\left(z^2+4z+1\right)\text{Li}_2(z)-\frac{1}{3}(2z+1)\left(z^2-8z+1\right)\zeta_2+\frac{1}{4}z(z+2)\log^2z\\
\nonumber&-\frac{1}{4}z(5z+8)\log z-\frac{1}{216}(1-z)\left(448z^2-1625z-281\right)\Big\}\\
\nonumber&+\epsilon ^3\Big\{\Big[\frac{64}{3}z\log z+\frac{16}{9}(1-z)\left(z^2-5z-2\right)\Big] \log^3(1-z) + \Big[32z\text{Li}_2(z)-32z\zeta_2\\
\nonumber&+4z\log^2z+4(2z+1)\log z-\frac{10}{9}(1-z)\left(4z^2-17z-5\right)\Big] \log^2(1-z)\\
\nonumber& + \Big[64z\text{Li}_3(1-z)+32z\text{Li}_3(z)-8z\text{Li}_2(z)\log z-32z\zeta_3-24z\zeta_2\log z\\
\nonumber&-\frac{16}{3}z\log^3z+4\left(z^2+4z+1\right)\text{Li}_2(z)+\frac{1}{2}\left(1-z^2\right)\log^2z\\
\nonumber&+\frac{4}{3}(2z+1)\left(z^2-8z+1\right)\zeta_2-\frac{5}{2}\left(1-z^2\right)\log z\\
\nonumber&+\frac{1}{54}(1-z)\left(448z^2-1625z-281\right)\Big] \log(1-z)-16z\zeta_2\text{Li}_2(z)\\
\nonumber&+32z\text{Li}_4\left(1-\frac{1}{z}\right)-32z\text{Li}_4(1-z)-28z\text{Li}_4(z)-z\text{Li}_2(z)\log^2z\\
\nonumber&-8z\text{Li}_3(1-z)\log z+2z\text{Li}_3(z)\log z+\frac{136}{5}z\zeta_2^2+15z\zeta_2\log^2z+10z\zeta_3\log z\\
\nonumber&+\frac{31}{24}z\log^4z+4\left(z^2+4z+1\right)\text{Li}_3(1-z)+\left(5z^2+8z-1\right)\text{Li}_3(z)\\
\nonumber&-\left(2z^2+2z-1\right)\text{Li}_2(z)\log z+\frac{1}{3}\left(-8z^3+33z^2-48z-13\right)\zeta_3-z(z+2)\zeta_2\log z\\
\nonumber&-\frac{1}{12}z(z+2)\log^3z+\frac{1}{2}\left(-5z^2-16z-5\right)\text{Li}_2(z)\\
\nonumber&+\frac{1}{18}\left(-40z^3+255z^2+24z-5\right)\zeta_2+\frac{1}{8}z(5z+8)\log^2z-\frac{9}{8}z(3z+4)\log z\\
\nonumber&-\frac{1}{1296}(1-z)\left(10496z^2-33385z-1897\right)\Big\}+\ord(\eps^4)\,,
\end{align}

\begin{align}
\XThree^H(z,\eps)&= -\log z \log(1-z)-\text{Li}_2(z)+\frac{\log^2z}{2}+\zeta_2+z-1+\epsilon\Big\{4\log z \log^2(1-z)\\
\nonumber& + \Big[4\text{Li}_2(z)-\frac{1}{2}\log^2z-4\zeta_2+2\log z+4(1-z)\Big] \log(1-z) + 6\text{Li}_3(1-z)\\
\nonumber&+7\text{Li}_3(z)-2\text{Li}_2(z)\log z-5\zeta_2\log z-\frac{1}{3}\log^3z-7\zeta_3+2\text{Li}_2(z)-\log^2z\\
\nonumber&-2\zeta_2-4(1-z)\Big\}+\epsilon ^2\Big\{-8\log z \log^3(1-z) + \Big[-8\text{Li}_2(z)-5\log^2z+8\zeta_2\\
\nonumber&-8\log z-8(1-z)\Big] \log^2(1-z) + \Big[-24\text{Li}_3(1-z)-28\text{Li}_3(z)+2\text{Li}_2(z)\log z\\
\nonumber&+30\zeta_2\log z+\frac{25}{6}\log^3z+28\zeta_3-8\text{Li}_2(z)+\log^2z+8\zeta_2+16(1-z)\Big] \log(1-z)\\
\nonumber& + 10\zeta_2\text{Li}_2(z)-3\text{Li}_2(z)^2-22\text{Li}_4\left(1-\frac{1}{z}\right)+6\text{Li}_4(1-z)+5\text{Li}_4(z)+\text{Li}_2(z)\log^2z\\
\nonumber&+2\text{Li}_3(z)\log z-\frac{15}{2}\zeta_2\log^2z-7\zeta_3\log z-\frac{19}{24}\log^4z-9\zeta_2^2-12\text{Li}_3(1-z)\\
\nonumber&-14\text{Li}_3(z)+4\text{Li}_2(z)\log z+10\zeta_2\log z+\frac{2}{3}\log^3z+14\zeta_3+4(1-z)\zeta_2\\
\nonumber&-16(1-z)\Big\}+\ord(\eps^3)\,,
\end{align}

\begin{align}
\XFour^H(z,\eps)&=\frac{1}{\epsilon ^2}\Big[-\log z+z-1\Big]+\frac{1}{\epsilon }\Big[4(1-z) \log(1-z)-4\text{Li}_2(z)+\log^2z+4\zeta_2\\
\nonumber&+4\log z\Big]+\Big[8\log z-8(1-z)\Big] \log^2(1-z) + \Big[16\text{Li}_2(z)-16\zeta_2\Big] \log(1-z)\\
\nonumber& + 16\text{Li}_3(1-z)+12\text{Li}_3(z)-2\text{Li}_2(z)\log z-6\zeta_2\log z-\frac{2}{3}\log^3z-12\zeta_3\\
\nonumber&+16\text{Li}_2(z)-4(z+3)\zeta_2-4\log^2z-4\log z-4(1-z)\\
\nonumber&+\epsilon\Big\{\Big[-\frac{64}{3}\log z+\frac{32}{3}(1-z)\Big] \log^3(1-z) + \Big[-32\text{Li}_2(z)-8\log^2z+32\zeta_2\\
\nonumber&-32\log z\Big] \log^2(1-z) + \Big[-64\text{Li}_3(1-z)-48\text{Li}_3(z)+8\text{Li}_2(z)\log z+40\zeta_2\log z\\
\nonumber&+8\log^3z+48\zeta_3+16(z+3)\zeta_2-64\text{Li}_2(z)+16(1-z)\Big] \log(1-z) + 16\zeta_2\text{Li}_2(z)\\
\nonumber&-48\text{Li}_4\left(1-\frac{1}{z}\right)+16\text{Li}_4(1-z)+26\text{Li}_4(z)+\text{Li}_2(z)\log^2z+8\text{Li}_3(1-z)\log z\\
\nonumber&-18\zeta_2\log^2z-10\zeta_3\log z-\frac{5}{3}\log^4z-\frac{132}{5}\zeta_2^2-64\text{Li}_3(1-z)-48\text{Li}_3(z)\\
\nonumber&+8\text{Li}_2(z)\log z-16(z-4)\zeta_3+24\zeta_2\log z+\frac{8}{3}\log^3z-16\text{Li}_2(z)+4\log^2z\\
\nonumber&+16\zeta_2-16(1-z)\Big\}+\ord(\eps^2)\,,
\end{align}

\begin{align}
\XFive^H(z,\eps)&=\frac{(1-z)^{-4\epsilon}}{\epsilon^2}\Big[2-4\epsilon+(2-8\zeta_2)\epsilon^2+(16\zeta_2-32\zeta_3)\epsilon^3+\ord(\eps^4)\Big]\\
\nonumber&
+\frac{1}{(1-z) \epsilon ^2}\Big[-2\log z-2(1-z)\Big]+\frac{1}{(1-z) \epsilon }\Big\{\Big[4\log z+8(1-z)\Big] \log(1-z)\\
\nonumber& + 4\zeta_2-4\text{Li}_2(z)+8\log z+4(1-z)\Big\}+\frac{1}{1-z}\Big\{-16(1-z) \log^2(1-z)\\
\nonumber& + \Big[16\text{Li}_2(z)-\log^2z-16\zeta_2-16\log z-16(1-z)\Big] \log(1-z) + 10\text{Li}_3(1-z)\\
\nonumber&+6\text{Li}_3(z)-4\text{Li}_2(z)\log z+6\zeta_2\log z+\frac{\log^3z}{3}-6\zeta_3+16\text{Li}_2(z)-8(z+1)\zeta_2\\
\nonumber&-8\log z-2(1-z)\Big\}+\frac{\epsilon }{1-z}\Big\{\Big[-\frac{32}{3}\log z+\frac{64}{3}(1-z)\Big] \log^3(1-z)\\
\nonumber& + \Big[-32\text{Li}_2(z)-\log^2z+32\zeta_2+32(1-z)\Big] \log^2(1-z) + \Big[-40\text{Li}_3(1-z)\\
\nonumber&-24\text{Li}_3(z)+6\text{Li}_2(z)\log z+2\zeta_2\log z-\log^3z+24\zeta_3-64\text{Li}_2(z)+32(z+1)\zeta_2\\
\nonumber&+4\log^2z+16\log z+8(1-z)\Big] \log(1-z) + 26\zeta_2\text{Li}_2(z)-5\text{Li}_2(z)^2\\
\nonumber&+6\text{Li}_4\left(1-\frac{1}{z}\right)+30\text{Li}_4(1-z)+26\text{Li}_4(z)+3\text{Li}_2(z)\log^2z-10\text{Li}_3(1-z)\log z\\
\nonumber&-14\text{Li}_3(z)\log z-2\zeta_2\log^2z+20\zeta_3\log z+\frac{\log^4z}{12}-\frac{157}{5}\zeta_2^2-40\text{Li}_3(1-z)\\
\nonumber&-24\text{Li}_3(z)+16\text{Li}_2(z)\log z-8(4z-7)\zeta_3-24\zeta_2\log z-\frac{4}{3}\log^3z+16z\zeta_2\\
\nonumber&-16\text{Li}_2(z)\Big\}+\ord(\eps^2)\,,
\end{align}

\begin{align}
\XSix^H(z,\eps)&=\frac{(1-z)^{-4\epsilon}}{\epsilon^3}\Big[-1+3\epsilon+(4\zeta_2-4)\epsilon^2+(-12\zeta_2+16\zeta_3+2)\epsilon^3\\
\nonumber&+\Big(\frac{72}{5}\zeta_2^2+16\zeta_2-48\zeta_3\Big)\epsilon^4+\ord(\eps^5)\Big]\\
\nonumber&
-\frac{1-z}{ z \epsilon ^3}+\frac{1}{(1-z) z \epsilon ^2}\Big[4(1-z)^2 \log(1-z) + \log z-(1-z)(3z-4)\Big]\\
\nonumber&+\frac{1}{(1-z) z \epsilon }\Big\{-8(1-z)^2 \log^2(1-z) + 4(1-z)(3z-4) \log(1-z) + 4\text{Li}_2(z)\\
\nonumber&+4(z-2)z\zeta_2-4\log z-4(1-z)^2\Big\}+\frac{1}{(1-z) z}\Big\{\frac{32}{3}(1-z)^2 \log^3(1-z) \\
\nonumber&+ \Big[-8\log z-8(1-z)(3z-4)\Big] \log^2(1-z) + \Big[-16\text{Li}_2(z)-16(z-2)z\zeta_2\\
\nonumber&+16(1-z)^2\Big] \log(1-z)-10\text{Li}_3(1-z)+4\text{Li}_3(z)-2\text{Li}_2(z)\log z\\
\nonumber&+4(2z-3)(2z-1)\zeta_3-6\zeta_2\log z-\frac{2}{3}\log^3z-16\text{Li}_2(z)-4z(3z-7)\zeta_2+4\log z\\
\nonumber&-2(1-z)z\Big\}+\frac{\epsilon }{(1-z) z}\Big\{-\frac{32}{3}(1-z)^2 \log^4(1-z) + \Big[\frac{64}{3}\log z\\
\nonumber&+\frac{32}{3}(1-z)(3z-4)\Big] \log^3(1-z) + \Big[32\text{Li}_2(z)+32(z-2)z\zeta_2-9\log^2z+32\log z\\
\nonumber&-32(1-z)^2\Big] \log^2(1-z) + \Big[40\text{Li}_3(1-z)-16\text{Li}_3(z)-10\text{Li}_2(z)\log z\\
\nonumber&-16(2z-3)(2z-1)\zeta_3+26\zeta_2\log z+\frac{4}{3}\log^3z+64\text{Li}_2(z)+16z(3z-7)\zeta_2\\
\nonumber&+8(1-z)z\Big] \log(1-z) + 2\zeta_2\text{Li}_2(z)-9\text{Li}_2(z)^2-2\text{Li}_4\left(1-\frac{1}{z}\right)-26\text{Li}_4(1-z)\\
\nonumber&-26\text{Li}_4(z)+4\text{Li}_2(z)\log^2z-18\text{Li}_3(1-z)\log z-2\text{Li}_3(z)\log z\\
\nonumber&+\frac{3}{5}\left(24z^2-48z+53\right)\zeta_2^2+11\zeta_2\log^2z+12\zeta_3\log z+\frac{7}{12}\log^4z+40\text{Li}_3(1-z)\\
\nonumber&-16\text{Li}_3(z)+8\text{Li}_2(z)\log z-16\left(3z^2-7z+3\right)\zeta_3+24\zeta_2\log z+\frac{8}{3}\log^3z+16\text{Li}_2(z)\\
\nonumber&+16(z-2)z\zeta_2\Big\}+\ord(\eps^2)\,,
\end{align}

\begin{align}
\XEight^H(z,\eps)&=
\frac{(1-z)^{-4\epsilon}}{\epsilon^2}\Big[2-8\epsilon+(8-8\zeta_2)\epsilon^2+(32\zeta_2-32\zeta_3)\epsilon^3+\ord(\eps^4)\Big]\\
\nonumber&+\frac{1}{(1-z) \epsilon ^2}\Big[-2\log z-2(1-z)\Big]+\frac{1}{(1-z) \epsilon }\Big\{\Big[8\log z+8(1-z)\Big] \log(1-z)\\
\nonumber& + 2\log^2z+8\log z+8(1-z)\Big\}+\frac{1}{1-z}\Big\{\Big[-16\log z-16(1-z)\Big] \log^2(1-z)\\
\nonumber& + \Big[-6\log^2z-32\log z-32(1-z)\Big] \log(1-z) + 36\text{Li}_3(z)-16\text{Li}_2(z)\log z\\
\nonumber&-12\zeta_2\log z-\frac{4}{3}\log^3z-36\zeta_3+8(1-z)\zeta_2-8\log^2z-8\log z-8(1-z)\Big\}\\
\nonumber&+\frac{\epsilon }{1-z}\Big\{\Big[\frac{64}{3}\log z+\frac{64}{3}(1-z)\Big] \log^3(1-z) + \Big[-24\log^2z+64\log z\\
\nonumber&+64(1-z)\Big] \log^2(1-z) + \Big[-144\text{Li}_3(z)+112\zeta_2\log z+\frac{22}{3}\log^3z+144\zeta_3\\
\nonumber&+24\log^2z-32(1-z)\zeta_2+32\log z+32(1-z)\Big] \log(1-z) + 64\zeta_2\text{Li}_2(z)\\
\nonumber&-32\text{Li}_2(z)^2-24\text{Li}_4\left(1-\frac{1}{z}\right)-24\text{Li}_4(1-z)+12\text{Li}_2(z)\log^2z-56\text{Li}_3(1-z)\log z\\
\nonumber&-20\text{Li}_3(z)\log z+52\zeta_3\log z-\frac{1}{3}\log^4z-32\zeta_2^2-144\text{Li}_3(z)+64\text{Li}_2(z)\log z\\
\nonumber&-16(2z-11)\zeta_3+48\zeta_2\log z+\frac{16}{3}\log^3z+8\log^2z-32(1-z)\zeta_2\Big\}
+\ord(\eps^2)\,,
\end{align}

\begin{align}
\XTen^H(z,\eps)&=-2\text{Li}_2(-z)+\frac{1}{2}\log^2z-2\log(1+z)\log z-\zeta_2-\frac{1}{6}(1-z)^3\\
\nonumber&+\epsilon\Big\{\Big[8\text{Li}_2(-z)-4\log^2(1+z)+8\log z\log(1+z)+8\log2\log(1+z)+4\zeta_2\\
\nonumber&-4\log^22+\frac{2}{3}(1-z)^3\Big] \log(1-z)-8\text{Li}_3\left(\frac{1-z}{2}\right)+8\text{Li}_3(1-z)+3\text{Li}_3(-z)\\
\nonumber&+4\text{Li}_3(z)+6\text{Li}_3\left(\frac{1}{1+z}\right)-8\text{Li}_3\left(\frac{1-z}{1+z}\right)-8\text{Li}_3\left(\frac{1+z}{2}\right)-\text{Li}_2(-z)\log z\\
\nonumber&-3\zeta_2\log z+11\zeta_2\log(1+z)-\frac{1}{3}\log^3z+\frac{1}{3}\log^3(1+z)+\frac{1}{2}\log(1+z)\log^2z\\
\nonumber&-4\log^22\log(1+z)+\zeta_3-8\zeta_2\log2+\frac{8}{3}\log^32+4\text{Li}_2(-z)-\log^2z\\
\nonumber&+4\log(1+z)\log z+2\zeta_2-\frac{5}{9}(1-z)^3\Big\}+\epsilon ^2\Big\{\Big[-16\text{Li}_2(-z)-4\log^2z\\
\nonumber&+16\log^2(1+z)-16\log(1+z)\log z-32\log2\log(1+z)-8\zeta_2+16\log^22\\
\nonumber&-\frac{4}{3}(1-z)^3\Big] \log^2(1-z) + \Big[32\text{Li}_3\left(\frac{1-z}{2}\right)-32\text{Li}_3(1-z)-12\text{Li}_3(-z)\\
\nonumber&-16\text{Li}_3(z)-24\text{Li}_3\left(\frac{1}{1+z}\right)+32\text{Li}_3\left(\frac{1-z}{1+z}\right)+32\text{Li}_3\left(\frac{1+z}{2}\right)\\
\nonumber&+4\text{Li}_2(-z)\log z+12\zeta_2\log z-44\zeta_2\log(1+z)+\frac{14}{3}\log^3z-\frac{4}{3}\log^3(1+z)\\
\nonumber&-2\log(1+z)\log^2z-2\log^2(1+z)\log z-2\log^22\log z+16\log^22\log(1+z)\\
\nonumber&+4\log2\log(1+z)\log z-4\zeta_3+32\zeta_2\log2-\frac{32}{3}\log^32-16\text{Li}_2(-z)\\
\nonumber&+8\log^2(1+z)-16\log z\log(1+z)-16\log2\log(1+z)-8\zeta_2+8\log^22\\
\nonumber&+\frac{20}{9}(1-z)^3\Big] \log(1-z)-3\text{Li}_4\left(1-\frac{1}{z^2}\right)+3\text{Li}_4\left(1-z^2\right)+8\zeta_2\text{Li}_2(-z)\\
\nonumber&-4\text{Li}_4\left(1-\frac{1}{z}\right)+4\text{Li}_4(1-z)+6\text{Li}_4(-z)+14\text{Li}_4(z)+15\text{Li}_4\left(\frac{1}{1+z}\right)\\
\nonumber&-20\text{Li}_4\left(\frac{1-z}{1+z}\right)+20\text{Li}_4\left(\frac{z-1}{z+1}\right)-15\text{Li}_4\left(\frac{z}{z+1}\right)+\frac{1}{2}\text{Li}_2(-z)\log^2z\\
\nonumber&-4\text{Li}_3\left(\frac{1-z}{2}\right)\log z+4\text{Li}_3(1-z)\log z+2\text{Li}_3(z)\log z+3\text{Li}_3\left(\frac{1}{1+z}\right)\log z\\
\nonumber&-4\text{Li}_3\left(\frac{1-z}{1+z}\right)\log z-4\text{Li}_3\left(\frac{1+z}{2}\right)\log z-\frac{11}{2}\zeta_2\log^2z+6\zeta_2\log(1+z)\log z\\
\nonumber&+3\zeta_3\log z-4\zeta_2\log2\log z-\frac{49}{24}\log^4z+4\log(1+z)\log^3z+\frac{8}{3}\log^3(1+z)\log z\\
\nonumber&+\frac{4}{3}\log^32\log z-\frac{15}{4}\log^2(1+z)\log^2z-2\log^22\log(1+z)\log z+\frac{1}{2}\zeta_2^2\\
\nonumber&+16\text{Li}_3\left(\frac{1-z}{2}\right)-16\text{Li}_3(1-z)-6\text{Li}_3(-z)-8\text{Li}_3(z)-12\text{Li}_3\left(\frac{1}{1+z}\right)\\
\nonumber&+16\text{Li}_3\left(\frac{1-z}{1+z}\right)+16\text{Li}_3\left(\frac{1+z}{2}\right)+2\text{Li}_2(-z)\log z+6\zeta_2\log z-22\zeta_2\log(1+z)\\
\nonumber&+\frac{2}{3}\log^3z-\frac{2}{3}\log^3(1+z)-\log(1+z)\log^2z+8\log^22\log(1+z)-2\zeta_3\\
\nonumber&+16\zeta_2\log2-\frac{16\log^32}{3}+\frac{2}{3}(1-z)^3\zeta_2-\frac{56}{27}(1-z)^3\Big\}+\ord(\eps^3)\,,
\end{align}

\begin{align}
\XEleven^H(z,\eps)&=\frac{1}{(z+1) \epsilon ^2}\Big[-2\log z+z^2-1\Big]+\frac{1}{(z+1) \epsilon }\Big[4\left(1-z^2\right) \log(1-z) \\
\nonumber&+ 4\text{Li}_2(-z)-8\text{Li}_2(z)-\log^2z+4\log(1+z)\log z+10\zeta_2+8\log z\Big]\\
\nonumber&+\frac{1}{z+1}\Big\{\Big[16\log z-8\left(1-z^2\right)\Big] \log^2(1-z) + \Big[-16\text{Li}_2(-z)+32\text{Li}_2(z)\\
\nonumber&+8\log^2(1+z)-16\log z\log(1+z)-16\log2\log(1+z)-40\zeta_2\\
\nonumber&+8\log^22\Big] \log(1-z) + 16\text{Li}_3\left(\frac{1-z}{2}\right)+16\text{Li}_3(1-z)-12\text{Li}_3(-z)\\
\nonumber&-8\text{Li}_3\left(\frac{1}{1+z}\right)+16\text{Li}_3\left(\frac{1-z}{1+z}\right)+16\text{Li}_3\left(\frac{1+z}{2}\right)+8\text{Li}_2(-z)\log z\\
\nonumber&-4\text{Li}_2(z)\log z+10\zeta_2\log z-20\zeta_2\log(1+z)+\frac{4}{3}\log^3z-\frac{4}{3}\log^3(1+z)\\
\nonumber&+2\log(1+z)\log^2z+8\log^22\log(1+z)-18\zeta_3+16\zeta_2\log2-\frac{16}{3}\log^32\\
\nonumber&-16\text{Li}_2(-z)+32\text{Li}_2(z)-4\left(z^2+9\right)\zeta_2+4\log^2z-16\log(1+z)\log z-8\log z\\
\nonumber&-4\left(1-z^2\right)\Big\}+\frac{\epsilon }{z+1}\Big\{\Big[-\frac{128}{3}\log z+\frac{32}{3}\left(1-z^2\right)\Big] \log^3(1-z) + \Big[32\text{Li}_2(-z)\\
\nonumber&-64\text{Li}_2(z)+8\log^2z-32\log^2(1+z)+32\log(1+z)\log z+64\log2\log(1+z)\\
\nonumber&+80\zeta_2-32\log^22-64\log z\Big] \log^2(1-z) + \Big[-64\text{Li}_3\left(\frac{1-z}{2}\right)-64\text{Li}_3(1-z)\\
\nonumber&+48\text{Li}_3(-z)+32\text{Li}_3\left(\frac{1}{1+z}\right)-64\text{Li}_3\left(\frac{1-z}{1+z}\right)-64\text{Li}_3\left(\frac{1+z}{2}\right)\\
\nonumber&-8\text{Li}_2(-z)\log z+16\text{Li}_2(z)\log z+4\zeta_2\log z+80\zeta_2\log(1+z)-8\log^3z\\
\nonumber&+\frac{16}{3}\log^3(1+z)+16\log(1+z)\log^2z+4\log^2(1+z)\log z+4\log^22\log z\\
\nonumber&-32\log^22\log(1+z)-8\log2\log(1+z)\log z+72\zeta_3-64\zeta_2\log2+\frac{64}{3}\log^32\\
\nonumber&+64\text{Li}_2(-z)-128\text{Li}_2(z)+16\left(z^2+9\right)\zeta_2-32\log^2(1+z)+64\log z\log(1+z)\\
\nonumber&+64\log2\log(1+z)-32\log^22+16\left(1-z^2\right)\Big] \log(1-z) + 12\text{Li}_4\left(1-\frac{1}{z^2}\right)\\
\nonumber&+4\text{Li}_4\left(1-z^2\right)-38\zeta_2\text{Li}_2(-z)+32\zeta_2\text{Li}_2(z)+2\text{Li}_2(-z)^2+24\text{Li}_2(-z)\text{Li}_2(z)\\
\nonumber&-48\text{Li}_4\left(1-\frac{1}{z}\right)+48\text{Li}_4(1-z)-34\text{Li}_4(-z)+8\text{Li}_4(z)+24\text{Li}_4\left(\frac{1}{1+z}\right)\\
\nonumber&+48\text{Li}_4\left(\frac{1-z}{1+z}\right)-48\text{Li}_4\left(\frac{z-1}{z+1}\right)+72\text{Li}_4\left(\frac{z}{z+1}\right)-9\text{Li}_2(-z)\log^2z\\
\nonumber&+2\text{Li}_2(z)\log^2z+4\text{Li}_2(-z)\log(1+z)\log z+8\text{Li}_3\left(\frac{1-z}{2}\right)\log z\\
\nonumber&+8\text{Li}_3(1-z)\log z+16\text{Li}_3(-z)\log z-12\text{Li}_3(z)\log z+12\text{Li}_3\left(\frac{1}{1+z}\right)\log z\\
\nonumber&+8\text{Li}_3\left(\frac{1-z}{1+z}\right)\log z+8\text{Li}_3\left(\frac{1+z}{2}\right)\log z+48\text{Li}_3(z)\log(1+z)-13\zeta_2\log^2z\\
\nonumber&-24\zeta_2\log^2(1+z)-28\zeta_2\log(1+z)\log z-44\zeta_3\log z+8\zeta_2\log2\log z\\
\nonumber&+36\zeta_3\log(1+z)+\frac{21}{4}\log^4z+4\log^4(1+z)-\frac{56}{3}\log(1+z)\log^3z\\
\nonumber&-\frac{46}{3}\log^3(1+z)\log z-\frac{8}{3}\log^32\log z+20\log^2(1+z)\log^2z\\
\nonumber&+4\log^22\log(1+z)\log z-\frac{486}{5}\zeta_2^2-64\text{Li}_3\left(\frac{1-z}{2}\right)-64\text{Li}_3(1-z)+48\text{Li}_3(-z)\\
\nonumber&+32\text{Li}_3\left(\frac{1}{1+z}\right)-64\text{Li}_3\left(\frac{1-z}{1+z}\right)-64\text{Li}_3\left(\frac{1+z}{2}\right)-32\text{Li}_2(-z)\log z\\
\nonumber&+16\text{Li}_2(z)\log z-8\left(2z^2-11\right)\zeta_3-40\zeta_2\log z+80\zeta_2\log(1+z)-\frac{16}{3}\log^3z\\
\nonumber&+\frac{16}{3}\log^3(1+z)-8\log(1+z)\log^2z-32\log^22\log(1+z)-64\zeta_2\log2+\frac{64}{3}\log^32\\
\nonumber&+16\text{Li}_2(-z)-32\text{Li}_2(z)-4\log^2z+16\log(1+z)\log z+40\zeta_2-16\left(1-z^2\right)\Big\}\\
\nonumber&+\ord(\eps^2)\,,
\end{align}

\begin{align}
\XTwelve^H(z,\eps)&=
\frac{1}{\epsilon ^2}\Big[-\log z+z-1\Big]+\frac{1}{\epsilon }\Big[4(1-z) \log(1-z)-4\text{Li}_2(z)+\log^2z+4\zeta_2\\
\nonumber&+4\log z\Big]+\Big[8\log z-8(1-z)\Big] \log^2(1-z) + \Big[16\text{Li}_2(z)-16\zeta_2\Big] \log(1-z)\\
\nonumber& + 16\text{Li}_3(1-z)+12\text{Li}_3(z)-2\text{Li}_2(z)\log z-6\zeta_2\log z-\frac{2}{3}\log^3z-12\zeta_3\\
\nonumber&+16\text{Li}_2(z)-4(z+3)\zeta_2-4\log^2z-4\log z-4(1-z)\\
\nonumber&+\epsilon\Big\{\Big[-\frac{64}{3}\log z+\frac{32}{3}(1-z)\Big] \log^3(1-z) + \Big[-32\text{Li}_2(z)-8\log^2z+32\zeta_2\\
\nonumber&-32\log z\Big] \log^2(1-z) + \Big[-64\text{Li}_3(1-z)-48\text{Li}_3(z)+8\text{Li}_2(z)\log z\\
\nonumber&+40\zeta_2\log z+8\log^3z+48\zeta_3+16(z+3)\zeta_2-64\text{Li}_2(z)+16(1-z)\Big] \log(1-z) \\
\nonumber&+ 16\zeta_2\text{Li}_2(z)-48\text{Li}_4\left(1-\frac{1}{z}\right)+16\text{Li}_4(1-z)+26\text{Li}_4(z)+\text{Li}_2(z)\log^2z\\
\nonumber&+8\text{Li}_3(1-z)\log z-18\zeta_2\log^2z-10\zeta_3\log z-\frac{5}{3}\log^4z-\frac{132}{5}\zeta_2^2-64\text{Li}_3(1-z)\\
\nonumber&-48\text{Li}_3(z)+8\text{Li}_2(z)\log z-16(z-4)\zeta_3+24\zeta_2\log z+\frac{8}{3}\log^3z-16\text{Li}_2(z)\\
\nonumber&+4\log^2z+16\zeta_2-16(1-z)\Big\}
+\ord(\eps^2)\,,
\end{align}

\begin{align}
\XThirteen^H(z,\eps)&=\frac{1}{z \epsilon ^2}\Big[-\log z-(1-z)z\Big]+\frac{1}{z \epsilon }\Big[4(1-z)z \log(1-z)-4\text{Li}_2(-z)\\
\nonumber&-4\text{Li}_2(z)+2\log^2z-4\log(1+z)\log z+2\zeta_2+4\log z\Big]\\
\nonumber&+\frac{1}{z}\Big\{\Big[8\log z-8(1-z)z\Big] \log^2(1-z) + \Big[16\text{Li}_2(-z)+16\text{Li}_2(z)-8\log^2(1+z)\\
\nonumber&+16\log z\log(1+z)+16\log2\log(1+z)-8\zeta_2-8\log^22\Big] \log(1-z)\\
\nonumber&-16\text{Li}_3\left(\frac{1-z}{2}\right)+32\text{Li}_3(1-z)+14\text{Li}_3(-z)+20\text{Li}_3(z)+12\text{Li}_3\left(\frac{1}{1+z}\right)\\
\nonumber&-16\text{Li}_3\left(\frac{1-z}{1+z}\right)-16\text{Li}_3\left(\frac{1+z}{2}\right)-6\text{Li}_2(-z)\log z-2\text{Li}_2(z)\log z-10\zeta_2\log z\\
\nonumber&+22\zeta_2\log(1+z)-2\log^3z+\frac{2}{3}\log^3(1+z)+\log(1+z)\log^2z-8\log^22\log(1+z)\\
\nonumber&-4\zeta_3-16\zeta_2\log2+\frac{16}{3}\log^32+16\text{Li}_2(-z)+16\text{Li}_2(z)-4\left(z^2-z+2\right)\zeta_2\\
\nonumber&-8\log^2z+16\log(1+z)\log z-4\log z-4(1-z)z\Big\}+\frac{\epsilon }{z}\Big\{\Big[-\frac{64}{3}\log z\\
\nonumber&+\frac{32}{3}(1-z)z\Big] \log^3(1-z) + \Big[-32\text{Li}_2(-z)-32\text{Li}_2(z)-16\log^2z+32\log^2(1+z)\\
\nonumber&-32\log(1+z)\log z-64\log2\log(1+z)+16\zeta_2+32\log^22-32\log z\Big] \log^2(1-z) \\
\nonumber&+ \Big[64\text{Li}_3\left(\frac{1-z}{2}\right)-128\text{Li}_3(1-z)-56\text{Li}_3(-z)-80\text{Li}_3(z)-48\text{Li}_3\left(\frac{1}{1+z}\right)\\
\nonumber&+64\text{Li}_3\left(\frac{1-z}{1+z}\right)+64\text{Li}_3\left(\frac{1+z}{2}\right)+8\text{Li}_2(-z)\log z+8\text{Li}_2(z)\log z+48\zeta_2\log z\\
\nonumber&-88\zeta_2\log(1+z)+\frac{68}{3}\log^3z-\frac{8}{3}\log^3(1+z)-20\log(1+z)\log^2z\\
\nonumber&-4\log^2(1+z)\log z-4\log^22\log z+32\log^22\log(1+z)+8\log2\log(1+z)\log z\\
\nonumber&+16\zeta_3+64\zeta_2\log2-\frac{64}{3}\log^32-64\text{Li}_2(-z)-64\text{Li}_2(z)+16\left(z^2-z+2\right)\zeta_2\\
\nonumber&+32\log^2(1+z)-64\log z\log(1+z)-64\log2\log(1+z)+32\log^22\\
\nonumber&+16(1-z)z\Big] \log(1-z)-14\text{Li}_4\left(1-\frac{1}{z^2}\right)-2\text{Li}_4\left(1-z^2\right)+34\zeta_2\text{Li}_2(-z)\\
\nonumber&+16\zeta_2\text{Li}_2(z)+2\text{Li}_2(-z)^2-16\text{Li}_2(-z)\text{Li}_2(z)-24\text{Li}_4\left(1-\frac{1}{z}\right)+56\text{Li}_4(1-z)\\
\nonumber&+14\text{Li}_4(-z)+70\text{Li}_4(z)-2\text{Li}_4\left(\frac{1}{1+z}\right)-40\text{Li}_4\left(\frac{1-z}{1+z}\right)+40\text{Li}_4\left(\frac{z-1}{z+1}\right)\\
\nonumber&-62\text{Li}_4\left(\frac{z}{z+1}\right)+2\text{Li}_2(-z)\log^2z+\text{Li}_2(z)\log^2z+4\text{Li}_2(-z)\log(1+z)\log z\\
\nonumber&-8\text{Li}_3\left(\frac{1-z}{2}\right)\log z+16\text{Li}_3(1-z)\log z+8\text{Li}_3(-z)\log z+4\text{Li}_3(z)\log z\\
\nonumber&+10\text{Li}_3\left(\frac{1}{1+z}\right)\log z-8\text{Li}_3\left(\frac{1-z}{1+z}\right)\log z-8\text{Li}_3\left(\frac{1+z}{2}\right)\log z\\
\nonumber&-32\text{Li}_3(z)\log(1+z)-24\zeta_2\log^2z+16\zeta_2\log^2(1+z)+32\zeta_2\log(1+z)\log z\\
\nonumber&+2\zeta_3\log z-8\zeta_2\log2\log z-24\zeta_3\log(1+z)-9\log^4z-\frac{8}{3}\log^4(1+z)\\
\nonumber&+\frac{56}{3}\log(1+z)\log^3z+10\log^3(1+z)\log z+\frac{8}{3}\log^32\log z-\frac{27}{2}\log^2(1+z)\log^2z\\
\nonumber&-4\log^22\log(1+z)\log z-\frac{11}{5}\zeta_2^2+64\text{Li}_3\left(\frac{1-z}{2}\right)-128\text{Li}_3(1-z)-56\text{Li}_3(-z)\\
\nonumber&-80\text{Li}_3(z)-48\text{Li}_3\left(\frac{1}{1+z}\right)+64\text{Li}_3\left(\frac{1-z}{1+z}\right)+64\text{Li}_3\left(\frac{1+z}{2}\right)+24\text{Li}_2(-z)\log z\\
\nonumber&+8\text{Li}_2(z)\log z-16\left(z^2-z-1\right)\zeta_3+40\zeta_2\log z-88\zeta_2\log(1+z)+8\log^3z\\
\nonumber&-\frac{8}{3}\log^3(1+z)-4\log(1+z)\log^2z+32\log^22\log(1+z)+64\zeta_2\log2-\frac{64}{3}\log^32\\
\nonumber&-16\text{Li}_2(-z)-16\text{Li}_2(z)+8\log^2z-16\log(1+z)\log z+8\zeta_2-16(1-z)z\Big\}\\
\nonumber&+\ord(\eps^2)\,,
\end{align}

\begin{align}
\XFourteen^H(z,\eps)&=\epsilon\Big[-\frac{1}{6}(z-3)z^2\log z-\frac{1}{36}(1-z)\left(5z^2-22z+5\right)\Big]\\
\nonumber&+\epsilon ^2\Big\{\Big[-\frac{1}{3}(1-z)^3\log z+\frac{1}{9}(1-z)\left(5z^2-22z+5\right)\Big] \log(1-z)\\
\nonumber&-\frac{1}{3}(z+1)\left(z^2-4z+1\right)\text{Li}_2(z)+\frac{1}{3}(z+1)\left(z^2-4z+1\right)\zeta_2+\frac{1}{12}(z-3)z^2\log^2z\\
\nonumber&-\frac{1}{12}z\left(5z^2-7z+4\right)\log z-\frac{1}{216}(1-z)\left(205z^2-662z+205\right)\Big\}\\
\nonumber&+\epsilon ^3\Big\{\Big[-\frac{4}{3}(3z-1)\log z-\frac{2}{9}(1-z)\left(5z^2-22z+5\right)\Big] \log^2(1-z) \\
\nonumber&+ \Big[\frac{4}{3}(z+1)\left(z^2-4z+1\right)\text{Li}_2(z)-\frac{4}{3}(z+1)\left(z^2-4z+1\right)\zeta_2+\frac{1}{6}(1-z)^3\log^2z\\
\nonumber&-\frac{1}{6}(1-z)\left(5z^2-6z+5\right)\log z+\frac{1}{54}(1-z)\left(205z^2-662z+205\right)\Big] \log(1-z)\\
\nonumber& + \frac{4}{3}(z+1)\left(z^2-4z+1\right)\text{Li}_3(1-z)+\frac{1}{3}\left(5z^3-15z^2+3z-1\right)\text{Li}_3(z)\\
\nonumber&+\frac{1}{3}\left(-2z^3+6z^2-3z+1\right)\text{Li}_2(z)\log z-\frac{1}{3}(z-3)z^2\zeta_2\log z-\frac{1}{36}(z-3)z^2\log^3z\\
\nonumber&+\frac{1}{3}\left(-5z^3+15z^2-3z+1\right)\zeta_3-\frac{1}{6}(z+1)\left(5z^2-8z+5\right)\text{Li}_2(z)\\
\nonumber&+\frac{1}{24}z\left(5z^2-7z+4\right)\log^2z+\frac{1}{18}\left(5z^3+45z^2-63z+25\right)\zeta_2\\
\nonumber&-\frac{1}{24}z\left(27z^2-25z+28\right)\log z-\frac{1}{1296}(1-z)\left(6365z^2-17158z+6365\right)\Big\}\\
\nonumber&+\epsilon ^4\Big\{\Big[-\frac{8}{9}\left(z^3-3z^2-9z+3\right)\log z+\frac{8}{27}(1-z)\left(5z^2-22z+5\right)\Big] \log^3(1-z)\\
\nonumber& + \Big[-\frac{8}{3}(z+1)\left(z^2-4z+1\right)\text{Li}_2(z)+\frac{8}{3}(z+1)\left(z^2-4z+1\right)\zeta_2\\
\nonumber&+\frac{1}{3}\left(-3z^3+9z^2-3z+1\right)\log^2z+\frac{2}{3}\left(4z^2-7z+5\right)\log z\\
\nonumber&-\frac{1}{27}(1-z)\left(205z^2-662z+205\right)\Big] \log^2(1-z) \\
\nonumber&+ \Big[-\frac{16}{3}(z+1)\left(z^2-4z+1\right)\text{Li}_3(1-z)+\frac{2}{3}(z+1)\left(z^2-4z+1\right)\text{Li}_2(z)\log z\\
\nonumber&-\frac{4}{3}\left(5z^3-15z^2+3z-1\right)\text{Li}_3(z)+\frac{4}{3}\left(5z^3-15z^2+3z-1\right)\zeta_3\\
\nonumber&+\frac{2}{3}\left(7z^3-21z^2+3z-1\right)\zeta_2\log z+\frac{1}{18}\left(9z^3-27z^2-21z+7\right)\log^3z\\
\nonumber&+\frac{2}{3}(z+1)\left(5z^2-8z+5\right)\text{Li}_2(z)+\frac{1}{12}(1-z)\left(5z^2-6z+5\right)\log^2z\\
\nonumber&-\frac{2}{9}\left(5z^3+45z^2-63z+25\right)\zeta_2-\frac{1}{12}(1-z)\left(27z^2-26z+27\right)\log z\\
\nonumber&+\frac{1}{324}(1-z)\left(6365z^2-17158z+6365\right)\Big] \log(1-z)\\
\nonumber&-\frac{8}{3}(z+1)\left(z^2-4z+1\right)\text{Li}_4\left(1-\frac{1}{z}\right)+\frac{8}{3}(z+1)\left(z^2-4z+1\right)\text{Li}_4(1-z)\\
\nonumber&+\frac{7}{3}(z+1)\left(z^2-4z+1\right)\text{Li}_4(z)+\frac{2}{3}\left(5z^3-15z^2+3z-1\right)\zeta_2\text{Li}_2(z)\\
\nonumber&+\frac{1}{6}\left(2z^3-6z^2+3z-1\right)\text{Li}_2(z)\log^2z-\frac{4}{3}\left(z^3-3z^2+6z-2\right)\text{Li}_3(1-z)\log z\\
\nonumber&+\frac{1}{3}\left(-2z^3+6z^2-3z+1\right)\text{Li}_3(z)\log z+(1-z)^3\text{Li}_2(z)^2\\
\nonumber&+\frac{1}{15}\left(-49z^3+147z^2+57z-19\right)\zeta_2^2+\frac{1}{6}\left(-7z^3+21z^2+24z-8\right)\zeta_2\log^2z\\
\nonumber&+\frac{1}{3}\left(3z^3-9z^2+24z-8\right)\zeta_3\log z+\frac{1}{144}\left(-15z^3+45z^2+48z-16\right)\log^4z\\
\nonumber&+\frac{2}{3}(z+1)\left(5z^2-8z+5\right)\text{Li}_3(1-z)+\frac{1}{6}\left(25z^3-39z^2+27z-5\right)\text{Li}_3(z)\\
\nonumber&+\frac{1}{6}\left(-10z^3+18z^2-15z+5\right)\text{Li}_2(z)\log z-\frac{1}{6}z\left(5z^2-7z+4\right)\zeta_2\log z\\
\nonumber&-\frac{1}{72}z\left(5z^2-7z+4\right)\log^3z+\frac{1}{18}\left(-115z^3+333z^2-297z+55\right)\zeta_3\\
\nonumber&-\frac{1}{4}(z+1)\left(9z^2-8z+9\right)\text{Li}_2(z)+\frac{1}{48}z\left(27z^2-25z+28\right)\log^2z\\
\nonumber&+\frac{1}{108}\left(-167z^3+1761z^2-1707z+653\right)\zeta_2-\frac{1}{48}z\left(153z^2-131z+164\right)\log z\\
\nonumber&-\frac{5}{7776}(1-z)\left(35537z^2-86302z+35537\right)\Big\}+\ord(\eps^5)\,,
\end{align}

\begin{align}
\XFifteen^H(z,\eps)&=-(1-z)\log z \log(1-z)-(1-z)\text{Li}_2(z)+(1-z)\zeta_2+\frac{1}{2}(1-z)\log^2z\\
\nonumber&+2\log z-\frac{1}{12}(1-z)\left(z^2-2z-23\right)+\epsilon\Big\{4(1-z)\log z \log^2(1-z)\\
\nonumber& + \Big[4(1-z)\text{Li}_2(z)-4(1-z)\zeta_2+\frac{1}{2}(z-1)\log^2z+2(1-z)\log z\\
\nonumber&+\frac{1}{3}(1-z)\left(z^2-2z-23\right)\Big] \log(1-z) + 6(1-z)\text{Li}_3(1-z)+7(1-z)\text{Li}_3(z)\\
\nonumber&-2(1-z)\text{Li}_2(z)\log z-7(1-z)\zeta_3-5(1-z)\zeta_2\log z+\frac{1}{3}(z-1)\log^3z\\
\nonumber&-2(z-5)\text{Li}_2(z)+2(z-5)\zeta_2+(z-2)\log^2z-2z\log z\\
\nonumber&-\frac{1}{18}(1-z)\left(5z^2-10z-103\right)\Big\}+\epsilon ^2\Big\{-8(1-z)\log z \log^3(1-z) \\
\nonumber&+ \Big[-8(1-z)\text{Li}_2(z)+8(1-z)\zeta_2-5(1-z)\log^2z+8(z-3)\log z\\
\nonumber&-\frac{2}{3}(1-z)\left(z^2-2z-23\right)\Big] \log^2(1-z) + \Big[-24(1-z)\text{Li}_3(1-z)\\
\nonumber&-28(1-z)\text{Li}_3(z)+2(1-z)\text{Li}_2(z)\log z+28(1-z)\zeta_3+30(1-z)\zeta_2\log z\\
\nonumber&+\frac{25}{6}(1-z)\log^3z+8(z-5)\text{Li}_2(z)-8(z-5)\zeta_2+(1-z)\log^2z\\
\nonumber&-4(1-z)\log z+\frac{2}{9}(1-z)\left(5z^2-10z-103\right)\Big] \log(1-z)\\
\nonumber& + 10(1-z)\zeta_2\text{Li}_2(z)-3(1-z)\text{Li}_2(z)^2-22(1-z)\text{Li}_4\left(1-\frac{1}{z}\right)\\
\nonumber&+6(1-z)\text{Li}_4(1-z)+5(1-z)\text{Li}_4(z)+(1-z)\text{Li}_2(z)\log^2z\\
\nonumber&+2(1-z)\text{Li}_3(z)\log z-9(1-z)\zeta_2^2-\frac{15}{2}(1-z)\zeta_2\log^2z-7(1-z)\zeta_3\log z\\
\nonumber&-\frac{19}{24}(1-z)\log^4z+4(3z-11)\text{Li}_3(1-z)+2(7z-15)\text{Li}_3(z)\\
\nonumber&-4(z-2)\text{Li}_2(z)\log z-2(7z-15)\zeta_3-2(5z-7)\zeta_2\log z+\frac{1}{3}(3-2z)\log^3z\\
\nonumber&-4(z+1)\text{Li}_2(z)+\frac{1}{3}\left(-z^3+3z^2+33z-11\right)\zeta_2+z\log^2z-6z\log z\\
\nonumber&-\frac{2}{27}(1-z)\left(14z^2-28z-229\right)\Big\}+\ord(\eps^3)\,,
\end{align}

\begin{align}
\XSixteen^H(z,\eps)&=-2\text{Li}_3(z)+\text{Li}_2(z)\log z+\zeta_2\log z+\frac{1}{6}\log^3z+2\zeta_3+\frac{1}{2}(1-z)^2\\
\nonumber&+\epsilon\Big\{\log^2z \log^2(1-z) + \Big[8\text{Li}_3(z)-2\text{Li}_2(z)\log z-6\zeta_2\log z-\frac{4}{3}\log^3z-8\zeta_3\\
\nonumber&-2(1-z)^2\Big] \log(1-z)-2\zeta_2\text{Li}_2(z)+\text{Li}_2(z)^2+8\text{Li}_4\left(1-\frac{1}{z}\right)\\
\nonumber&+8\text{Li}_4(1-z)+8\text{Li}_4(z)+\frac{1}{2}\text{Li}_2(z)\log^2z-2\text{Li}_3(1-z)\log z-5\text{Li}_3(z)\log z\\
\nonumber&+\frac{1}{2}\zeta_2\log^2z-3\zeta_3\log z+\frac{1}{6}\log^4z-\frac{11}{5}\zeta_2^2+8\text{Li}_3(z)-4\text{Li}_2(z)\log z-4\zeta_2\log z\\
\nonumber&-\frac{2}{3}\log^3z-8\zeta_3+2(1-z)^2\Big\}+\ord(\eps^2)\,,
\end{align}

\begin{align}
\XSeventeen^H(z,\eps)&= \frac{1}{\epsilon }\Big[\frac{\log^2z}{2}-\frac{1}{2}(1-z)^2\Big]+2(1-z)^2 \log(1-z) + 4\text{Li}_3(z)-4\zeta_2\log z\\
\nonumber&-\frac{1}{2}\log^3z-4\zeta_3-2\log^2z-(1-z)^2+\epsilon\Big\{\Big[-4\log^2z-4(1-z)^2\Big] \log^2(1-z)\\
\nonumber& + \Big[-16\text{Li}_3(z)+16\zeta_2\log z+\frac{8}{3}\log^3z+16\zeta_3+4(1-z)^2\Big] \log(1-z)\\
\nonumber&-16\text{Li}_4\left(1-\frac{1}{z}\right)-16\text{Li}_4(1-z)-2\text{Li}_4(z)+2\text{Li}_3(z)\log z-3\zeta_2\log^2z\\
\nonumber&-\frac{3}{8}\log^4z+\frac{4}{5}\zeta_2^2-16\text{Li}_3(z)+16\zeta_2\log z+2\log^3z+16\zeta_3+2(1-z)^2\zeta_2\\
\nonumber&+2\log^2z-4(1-z)^2\Big\}+\ord(\eps^2)\,,
\end{align}

\begin{align}
\XEighteen^H(z,\eps)&= \frac{1}{(z+1) \epsilon ^2}\Big[-\log z+\frac{1}{2}\left(z^2-1\right)\Big]+\frac{1}{(z+1) \epsilon }\Big[2\left(1-z^2\right) \log(1-z)\\
\nonumber& + 4\text{Li}_2(-z)-4\text{Li}_2(z)-\log^2z+4\log(1+z)\log z+6\zeta_2+4\log z\Big]\\
\nonumber&+\frac{1}{z+1}\Big\{\Big[8\log z-4\left(1-z^2\right)\Big] \log^2(1-z) + \Big[-16\text{Li}_2(-z)+16\text{Li}_2(z)\\
\nonumber&+8\log^2(1+z)-16\log z\log(1+z)-16\log2\log(1+z)-24\zeta_2\\
\nonumber&+8\log^22\Big] \log(1-z) + 16\text{Li}_3\left(\frac{1-z}{2}\right)-8\text{Li}_3(-z)-8\text{Li}_3(z)+16\text{Li}_3\left(\frac{1-z}{1+z}\right)\\
\nonumber&+16\text{Li}_3\left(\frac{1+z}{2}\right)+8\text{Li}_2(-z)\log z+12\zeta_2\log z-16\zeta_2\log(1+z)+\log^3z\\
\nonumber&-\frac{8}{3}\log^3(1+z)+4\log(1+z)\log^2z+8\log^22\log(1+z)-14\zeta_3+16\zeta_2\log2\\
\nonumber&-\frac{16\log^32}{3}-16\text{Li}_2(-z)+16\text{Li}_2(z)-2\left(z^2+11\right)\zeta_2+4\log^2z-16\log(1+z)\log z\\
\nonumber&-4\log z-2\left(1-z^2\right)\Big\}+\frac{\epsilon }{z+1}\Big\{\Big[-\frac{64}{3}\log z+\frac{16}{3}\left(1-z^2\right)\Big] \log^3(1-z)\\
\nonumber& + \Big[32\text{Li}_2(-z)-32\text{Li}_2(z)+10\log^2z-32\log^2(1+z)+32\log(1+z)\log z\\
\nonumber&+64\log2\log(1+z)+48\zeta_2-32\log^22-32\log z\Big] \log^2(1-z) \\
\nonumber&+ \Big[-64\text{Li}_3\left(\frac{1-z}{2}\right)+32\text{Li}_3(-z)+32\text{Li}_3(z)-64\text{Li}_3\left(\frac{1-z}{1+z}\right)\\
\nonumber&-64\text{Li}_3\left(\frac{1+z}{2}\right)+4\text{Li}_2(z)\log z-20\zeta_2\log z+64\zeta_2\log(1+z)-\frac{32}{3}\log^3z\\
\nonumber&+\frac{32}{3}\log^3(1+z)+16\log(1+z)\log^2z-32\log^22\log(1+z)+56\zeta_3-64\zeta_2\log2\\
\nonumber&+\frac{64}{3}\log^32+64\text{Li}_2(-z)-64\text{Li}_2(z)+8\left(z^2+11\right)\zeta_2-32\log^2(1+z)\\
\nonumber&+64\log z\log(1+z)+64\log2\log(1+z)-32\log^22+8\left(1-z^2\right)\Big] \log(1-z) \\
\nonumber&+ 8\text{Li}_4\left(1-\frac{1}{z^2}\right)+8\text{Li}_4\left(1-z^2\right)-48\zeta_2\text{Li}_2(-z)+12\zeta_2\text{Li}_2(z)+2\text{Li}_2(z)^2\\
\nonumber&+32\text{Li}_2(-z)\text{Li}_2(z)-24\text{Li}_4(-z)-10\text{Li}_4(z)+64\text{Li}_4\left(\frac{1}{1+z}\right)+64\text{Li}_4\left(\frac{1-z}{1+z}\right)\\
\nonumber&-64\text{Li}_4\left(\frac{z-1}{z+1}\right)+64\text{Li}_4\left(\frac{z}{z+1}\right)-12\text{Li}_2(-z)\log^2z+2\text{Li}_2(z)\log^2z\\
\nonumber&+4\text{Li}_3(1-z)\log z+24\text{Li}_3(-z)\log z-14\text{Li}_3(z)\log z+32\text{Li}_3\left(\frac{1}{1+z}\right)\log z\\
\nonumber&+64\text{Li}_3(z)\log(1+z)+5\zeta_2\log^2z-32\zeta_2\log^2(1+z)-32\zeta_2\log(1+z)\log z\\
\nonumber&-24\zeta_3\log z+48\zeta_3\log(1+z)+\frac{29}{6}\log^4z+\frac{16}{3}\log^4(1+z)-\frac{44}{3}\log(1+z)\log^3z\\
\nonumber&-16\log^3(1+z)\log z+16\log^2(1+z)\log^2z-\frac{416\zeta_2^2}{5}-64\text{Li}_3\left(\frac{1-z}{2}\right)\\
\nonumber&+32\text{Li}_3(-z)+32\text{Li}_3(z)-64\text{Li}_3\left(\frac{1-z}{1+z}\right)-64\text{Li}_3\left(\frac{1+z}{2}\right)-32\text{Li}_2(-z)\log z\\
\nonumber&-8\left(z^2-8\right)\zeta_3-48\zeta_2\log z+64\zeta_2\log(1+z)-4\log^3z+\frac{32}{3}\log^3(1+z)\\
\nonumber&-16\log(1+z)\log^2z-32\log^22\log(1+z)-64\zeta_2\log2+\frac{64}{3}\log^32+16\text{Li}_2(-z)\\
\nonumber&-16\text{Li}_2(z)-4\log^2z+16\log(1+z)\log z+24\zeta_2-8\left(1-z^2\right)\Big\}+\ord(\eps^2)\,,
\end{align}

\begin{align}
\XTwenty^H(z,\eps)&=\frac{1-z}{2z \epsilon ^3}+\frac{1}{z (z+1) \epsilon ^2}\Big[-2\left(1-z^2\right) \log(1-z) + z\log z-2\left(1-z^2\right)\Big]\\
\nonumber&+\frac{1}{z (z+1) \epsilon }\Big[4\left(1-z^2\right) \log^2(1-z) + 8\left(1-z^2\right) \log(1-z) + 2(1-z)\text{Li}_2(-z)\\
\nonumber&+4z\text{Li}_2(z)+\left(2z^2-5z-1\right)\zeta_2-\log^2z+2(1-z)\log(1+z)\log z-4z\log z\\
\nonumber&+2\left(1-z^2\right)\Big]+\frac{1}{z (z+1)}\Big\{-\frac{16}{3}\left(1-z^2\right) \log^3(1-z) + \Big[-8z\log z\\
\nonumber&-16\left(1-z^2\right)\Big] \log^2(1-z) + \Big[-8(1-z)\text{Li}_2(-z)-16z\text{Li}_2(z)\\
\nonumber&-4\left(2z^2-5z-1\right)\zeta_2+(z+1)\log^2z+4(1-z)\log^2(1+z)+4(1-z)\log^22\\
\nonumber&-8(1-z)\log(1+z)\log z-8(1-z)\log2\log(1+z)-8\left(1-z^2\right)\Big] \log(1-z) \\
\nonumber&+ 8(1-z)\text{Li}_3\left(\frac{1-z}{2}\right)-10(z+1)\text{Li}_3(1-z)-4(z+3)\text{Li}_3(-z)\\
\nonumber&-2(3z+7)\text{Li}_3(z)+8(1-z)\text{Li}_3\left(\frac{1-z}{1+z}\right)+8(1-z)\text{Li}_3\left(\frac{1+z}{2}\right)\\
\nonumber&+8\text{Li}_2(-z)\log z+4(z+1)\text{Li}_2(z)\log z+\left(8z^2+11z-11\right)\zeta_3\\
\nonumber&-4(z-2)\zeta_2\log z-8(1-z)\zeta_2\log(1+z)+8(1-z)\zeta_2\log2+\frac{1}{3}(z+4)\log^3z\\
\nonumber&-\frac{4}{3}(1-z)\log^3(1+z)-\frac{8}{3}(1-z)\log^32+2(1-z)\log(1+z)\log^2z\\
\nonumber&+4(1-z)\log^22\log(1+z)-8(1-z)\text{Li}_2(-z)-16z\text{Li}_2(z)\\
\nonumber&-4\left(2z^2-5z-1\right)\zeta_2+4\log^2z-8(1-z)\log(1+z)\log z+4z\log z\Big\}\\
\nonumber&+\frac{\epsilon }{z (z+1)}\Big\{\frac{16}{3}\left(1-z^2\right) \log^4(1-z) + \Big[\frac{64}{3}z\log z+\frac{64}{3}\left(1-z^2\right)\Big] \log^3(1-z)\\
\nonumber& + \Big[16(1-z)\text{Li}_2(-z)+32z\text{Li}_2(z)+8\left(2z^2-5z-1\right)\zeta_2+(z+11)\log^2z\\
\nonumber&-16(1-z)\log^2(1+z)-16(1-z)\log^22+16(1-z)\log(1+z)\log z\\
\nonumber&+32(1-z)\log2\log(1+z)+32z\log z+16\left(1-z^2\right)\Big] \log^2(1-z) \\
\nonumber&+ \Big[-32(1-z)\text{Li}_3\left(\frac{1-z}{2}\right)+40(z+1)\text{Li}_3(1-z)+16(z+3)\text{Li}_3(-z)\\
\nonumber&+8(3z+7)\text{Li}_3(z)-32(1-z)\text{Li}_3\left(\frac{1-z}{1+z}\right)-32(1-z)\text{Li}_3\left(\frac{1+z}{2}\right)\\
\nonumber&-2(3z+1)\text{Li}_2(z)\log z-4\left(8z^2+11z-11\right)\zeta_3-10(z+3)\zeta_2\log z\\
\nonumber&+32(1-z)\zeta_2\log(1+z)-32(1-z)\zeta_2\log2+\frac{1}{3}(-13z-45)\log^3z\\
\nonumber&+\frac{16}{3}(1-z)\log^3(1+z)+\frac{32}{3}(1-z)\log^32+8(z+3)\log(1+z)\log^2z\\
\nonumber&-16(1-z)\log^22\log(1+z)+32(1-z)\text{Li}_2(-z)+64z\text{Li}_2(z)\\
\nonumber&+16\left(2z^2-5z-1\right)\zeta_2-4(z+1)\log^2z-16(1-z)\log^2(1+z)-16(1-z)\log^22\\
\nonumber&+32(1-z)\log(1+z)\log z+32(1-z)\log2\log(1+z)\Big] \log(1-z) \\
\nonumber&+ 4(z+3)\text{Li}_4\left(1-\frac{1}{z^2}\right)+4(z+3)\text{Li}_4\left(1-z^2\right)-48\zeta_2\text{Li}_2(-z)\\
\nonumber&-2(13z+7)\zeta_2\text{Li}_2(z)-8(z+1)\text{Li}_2(-z)^2+(5z+7)\text{Li}_2(z)^2+32\text{Li}_2(-z)\text{Li}_2(z)\\
\nonumber&-10(z+1)\text{Li}_4\left(1-\frac{1}{z}\right)-54(z+1)\text{Li}_4(1-z)+4(7z+1)\text{Li}_4(-z)\\
\nonumber&-2(9z+14)\text{Li}_4(z)+64\text{Li}_4\left(\frac{1}{1+z}\right)+32(1-z)\text{Li}_4\left(\frac{1-z}{1+z}\right)\\
\nonumber&-32(1-z)\text{Li}_4\left(\frac{z-1}{z+1}\right)+64\text{Li}_4\left(\frac{z}{z+1}\right)+2(5z-1)\text{Li}_2(-z)\log^2z\\
\nonumber&-(3z+1)\text{Li}_2(z)\log^2z-16(z+1)\text{Li}_2(-z)\log(1+z)\log z\\
\nonumber&+2(z+3)\text{Li}_3(1-z)\log z-12(3z+1)\text{Li}_3(-z)\log z+2(3z-4)\text{Li}_3(z)\log z\\
\nonumber&-32z\text{Li}_3\left(\frac{1}{1+z}\right)\log z+64\text{Li}_3(z)\log(1+z)+\frac{4}{5}\left(9z^2+50z-63\right)\zeta_2^2\\
\nonumber&+5\zeta_2\log^2z-32\zeta_2\log^2(1+z)-16(z+3)\zeta_2\log(1+z)\log z+6(3z-1)\zeta_3\log z\\
\nonumber&+48\zeta_3\log(1+z)+\frac{1}{12}(21z+79)\log^4z+\frac{16}{3}\log^4(1+z)\\
\nonumber&-\frac{2}{3}(5z+27)\log(1+z)\log^3z+\frac{16}{3}(z-2)\log^3(1+z)\log z\\
\nonumber&+8(1-z)\log^2(1+z)\log^2z-32(1-z)\text{Li}_3\left(\frac{1-z}{2}\right)+40(z+1)\text{Li}_3(1-z)\\
\nonumber&+16(z+3)\text{Li}_3(-z)+8(3z+7)\text{Li}_3(z)-32(1-z)\text{Li}_3\left(\frac{1-z}{1+z}\right)\\
\nonumber&-32(1-z)\text{Li}_3\left(\frac{1+z}{2}\right)-32\text{Li}_2(-z)\log z-16(z+1)\text{Li}_2(z)\log z\\
\nonumber&-4\left(8z^2+11z-11\right)\zeta_3+16(z-2)\zeta_2\log z+32(1-z)\zeta_2\log(1+z)\\
\nonumber&-32(1-z)\zeta_2\log2-\frac{4}{3}(z+4)\log^3z+\frac{16}{3}(1-z)\log^3(1+z)\\
\nonumber&+\frac{32}{3}(1-z)\log^32-8(1-z)\log(1+z)\log^2z-16(1-z)\log^22\log(1+z)\\
\nonumber&+8(1-z)\text{Li}_2(-z)+16z\text{Li}_2(z)+4\left(2z^2-5z-1\right)\zeta_2-4\log^2z\\
\nonumber&+8(1-z)\log(1+z)\log z\Big\}+\ord(\eps^2)\,,
\end{align}

\begin{align}
\XTwentyFive^H(z,\eps)&=\frac{(1-z)^{-4\epsilon}}{\epsilon^2}\Big[2-4\eps+(2-8\zeta_2)\epsilon^2+(16\zeta_2-32\zeta_3)\epsilon^3+\ord(\eps^4)\Big]\\
\nonumber&
+\frac{1}{(1-z) \epsilon ^2}\Big[-2\log z-2(1-z)\Big]+\frac{1}{(1-z) \epsilon }\Big\{\Big[4\log z+8(1-z)\Big] \log(1-z)\\
\nonumber& + 4\zeta_2-4\text{Li}_2(z)+8\log z+4(1-z)\Big\}+\frac{1}{1-z}\Big\{-16(1-z) \log^2(1-z)\\
\nonumber& + \Big[16\text{Li}_2(z)-\log^2z-16\zeta_2-16\log z-16(1-z)\Big] \log(1-z) + 10\text{Li}_3(1-z)\\
\nonumber&+6\text{Li}_3(z)-4\text{Li}_2(z)\log z+6\zeta_2\log z+\frac{1}{3}\log^3z-6\zeta_3+16\text{Li}_2(z)-8(z+1)\zeta_2\\
\nonumber&-8\log z-2(1-z)\Big\}+\frac{\epsilon }{1-z}\Big\{\Big[-\frac{32}{3}\log z+\frac{64(1-z)}{3}\Big] \log^3(1-z)\\
\nonumber& + \Big[-32\text{Li}_2(z)-\log^2z+32\zeta_2+32(1-z)\Big] \log^2(1-z) + \Big[-40\text{Li}_3(1-z)\\
\nonumber&-24\text{Li}_3(z)+6\text{Li}_2(z)\log z+2\zeta_2\log z-\log^3z+24\zeta_3-64\text{Li}_2(z)+32(z+1)\zeta_2\\
\nonumber&+4\log^2z+16\log z+8(1-z)\Big] \log(1-z) + 26\zeta_2\text{Li}_2(z)-5\text{Li}_2(z)^2\\
\nonumber&+6\text{Li}_4\left(1-\frac{1}{z}\right)+30\text{Li}_4(1-z)+26\text{Li}_4(z)+3\text{Li}_2(z)\log^2z-10\text{Li}_3(1-z)\log z\\
\nonumber&-14\text{Li}_3(z)\log z-2\zeta_2\log^2z+20\zeta_3\log z+\frac{1}{12}\log^4z-\frac{157}{5}\zeta_2^2-40\text{Li}_3(1-z)\\
\nonumber&-24\text{Li}_3(z)+16\text{Li}_2(z)\log z-8(4z-7)\zeta_3-24\zeta_2\log z-\frac{4}{3}\log^3z+16z\zeta_2\\
\nonumber&-16\text{Li}_2(z)\Big\}+\ord(\eps^2)\,,
\end{align}

\begin{align}
\XThirty^H(z,\eps)&=\frac{(1-z)^{-4\epsilon}}{\epsilon^2}\Big[1-4\eps+(4-4\zeta_2)\epsilon^2+(16\zeta_2-16\zeta_3)\epsilon^3+\ord(\eps^4)\Big]\\
\nonumber&+
\frac{1}{(1-z) \epsilon ^2}\Big[-\log z+z-1\Big]+\frac{1}{(1-z) \epsilon }\Big\{\Big[4\log z+4(1-z)\Big] \log(1-z)\\
\nonumber& + \log^2z+4\log z+4(1-z)\Big\}+\frac{1}{1-z}\Big\{\Big[-8\log z-8(1-z)\Big] \log^2(1-z)\\
\nonumber& + \Big[-4\log^2z-16\log z-16(1-z)\Big] \log(1-z) + 20\text{Li}_3(z)-10\text{Li}_2(z)\log z\\
\nonumber&-6\zeta_2\log z-\frac{2}{3}\log^3z-20\zeta_3+4(1-z)\zeta_2-4\log^2z-4\log z-4(1-z)\Big\}\\
\nonumber&+\frac{\epsilon }{1-z}\Big\{\Big[\frac{32}{3}\log z+\frac{32}{3}(1-z)\Big] \log^3(1-z) + \Big[-12\log^2z+32\log z\\
\nonumber&+32(1-z)\Big] \log^2(1-z) + \Big[-80\text{Li}_3(z)+64\zeta_2\log z+\frac{8}{3}\log^3z+80\zeta_3\\
\nonumber&+16\log^2z-16(1-z)\zeta_2+16\log z+16(1-z)\Big] \log(1-z) + 40\zeta_2\text{Li}_2(z)\\
\nonumber&-20\text{Li}_2(z)^2-6\text{Li}_4(z)+9\text{Li}_2(z)\log^2z-40\text{Li}_3(1-z)\log z-16\text{Li}_3(z)\log z\\
\nonumber&+6\zeta_2\log^2z+38\zeta_3\log z+\frac{1}{3}\log^4z-\frac{88}{5}\zeta_2^2-80\text{Li}_3(z)+40\text{Li}_2(z)\log z\\
\nonumber&-16(z-6)\zeta_3+24\zeta_2\log z+\frac{8}{3}\log^3z+4\log^2z-16(1-z)\zeta_2\Big\}+\ord(\eps^2)\,,
\end{align}


\pagebreak
\bibliographystyle{JHEP}

\begin{thebibliography}{100}

\bibitem{:2012gk}
  G.~Aad {\it et al.}  [ATLAS Collaboration],
  ``Observation of a new particle in the search for the Standard Model Higgs boson with the ATLAS detector at the LHC,''
  [arXiv:1207.7214 [hep-ex]].


\bibitem{:2012gu}
  S.~Chatrchyan {\it et al.}  [CMS Collaboration],
  ``Observation of a new boson at a mass of 125 GeV with the CMS experiment at the LHC,''
  [arXiv:1207.7235 [hep-ex]].


\bibitem{Chatrchyan:2011wt} 
  S.~Chatrchyan {\it et al.}  [CMS Collaboration],
  ``Measurement of the Rapidity and Transverse Momentum Distributions of Z Bosons in pp Collisions at sqrt(s)=7 TeV,''
  Phys.\ Rev.\ D {\bf 85}, 032002 (2012)
  [arXiv:1110.4973 [hep-ex]].

\bibitem{Aad:2011qv} 
  G.~Aad {\it et al.}  [ATLAS Collaboration],
  ``Measurement of the production cross section for Z/gamma* in association with jets in pp collisions at sqrt(s) = 7 TeV with the ATLAS detector,''
  Phys.\ Rev.\ D {\bf 85}, 032009 (2012)
  [arXiv:1111.2690 [hep-ex]].


\bibitem{Hamberg:1990np} 
  R.~Hamberg, W.~L.~van Neerven and T.~Matsuura,
  ``A Complete calculation of the order $\alpha_s^{2}$ correction to the Drell-Yan $K$ factor,''
  Nucl.\ Phys.\ B {\bf 359}, 343 (1991)
  [Erratum-ibid.\ B {\bf 644}, 403 (2002)].

\bibitem{Harlander:2002wh} 
  R.~V.~Harlander and W.~B.~Kilgore,
  ``Next-to-next-to-leading order Higgs production at hadron colliders,''
  Phys.\ Rev.\ Lett.\  {\bf 88}, 201801 (2002)
  [hep-ph/0201206].

\bibitem{Anastasiou:2002yz}
  C.~Anastasiou and K.~Melnikov,
  ``Higgs boson production at hadron colliders in NNLO QCD,''
  Nucl.\ Phys.\ B {\bf 646} (2002) 220
  [hep-ph/0207004].

\bibitem{Anastasiou:2002wq} 
  C.~Anastasiou and K.~Melnikov,
  ``Pseudoscalar Higgs boson production at hadron colliders in NNLO QCD,''
  Phys.\ Rev.\ D {\bf 67}, 037501 (2003)
  [hep-ph/0208115].

\bibitem{Harlander:2002vv} 
  R.~V.~Harlander and W.~B.~Kilgore,
  ``Production of a pseudoscalar Higgs boson at hadron colliders at next-to-next-to leading order,''
  JHEP {\bf 0210}, 017 (2002)
  [hep-ph/0208096].

\bibitem{Harlander:2003ai} 
  R.~V.~Harlander and W.~B.~Kilgore,
  ``Higgs boson production in bottom quark fusion at next-to-next-to leading order,''
  Phys.\ Rev.\ D {\bf 68}, 013001 (2003)
  [hep-ph/0304035].

\bibitem{Brein:2003wg} 
  O.~Brein, A.~Djouadi and R.~Harlander,
  ``NNLO QCD corrections to the Higgs-strahlung processes at hadron colliders,''
  Phys.\ Lett.\ B {\bf 579}, 149 (2004)
  [hep-ph/0307206].

\bibitem{Pak:2009dg} 
  A.~Pak, M.~Rogal and M.~Steinhauser,
  ``Finite top quark mass effects in NNLO Higgs boson production at LHC,''
  JHEP {\bf 1002}, 025 (2010)
  [arXiv:0911.4662 [hep-ph]].

\bibitem{Harlander:2009mq} 
  R.~V.~Harlander and K.~J.~Ozeren,
  ``Finite top mass effects for hadronic Higgs production at next-to-next-to-leading order,''
  JHEP {\bf 0911}, 088 (2009)
  [arXiv:0909.3420 [hep-ph]].
  
\bibitem{Pak:2011hs}
  A.~Pak, M.~Rogal and M.~Steinhauser,
  ``Production of scalar and pseudo-scalar Higgs bosons to next-to-next-to-leading order at hadron colliders,''
  JHEP {\bf 1109} (2011) 088
  [arXiv:1107.3391 [hep-ph]].

\bibitem{Ravindran:2003um} 
  V.~Ravindran, J.~Smith and W.~L.~van Neerven,
  ``NNLO corrections to the total cross-section for Higgs boson production in hadron hadron collisions,''
  Nucl.\ Phys.\ B {\bf 665}, 325 (2003)
  [hep-ph/0302135].


\bibitem{Anastasiou:2012hx} 
  C.~Anastasiou, S.~Buehler, F.~Herzog and A.~Lazopoulos,
  ``Inclusive Higgs boson cross-section for the LHC at 8 TeV,''
  JHEP {\bf 1204}, 004 (2012)
  [arXiv:1202.3638 [hep-ph]].

\bibitem{Anastasiou:2003ds} 
  C.~Anastasiou, L.~J.~Dixon, K.~Melnikov and F.~Petriello,
  ``High precision QCD at hadron colliders: Electroweak gauge boson rapidity distributions at NNLO,''
  Phys.\ Rev.\ D {\bf 69}, 094008 (2004)
  [hep-ph/0312266].
  
\bibitem{Gonsalves:1983nq}
  R.~J.~Gonsalves,
  ``Dimensionally Regularized Two Loop On-shell Quark Form-factor,''
  Phys.\ Rev.\ D {\bf 28} (1983) 1542.
  
\bibitem{Kramer:1986sr}
  G.~Kramer and B.~Lampe,
  ``Integrals For Two Loop Calculations In Massless Qcd,''
  J.\ Math.\ Phys.\  {\bf 28} (1987) 945.
  
\bibitem{Gehrmann:2005pd}
  T.~Gehrmann, T.~Huber and D.~Maitre,
  ``Two-loop quark and gluon form-factors in dimensional regularization,''
  Phys.\ Lett.\ B {\bf 622} (2005) 295
  [hep-ph/0507061].
  
\bibitem{Huber:2005yg} 
  T.~Huber and D.~Maitre,
  ``HypExp: A Mathematica package for expanding hypergeometric functions around integer-valued parameters,''
  Comput.\ Phys.\ Commun.\  {\bf 175}, 122 (2006)
  [hep-ph/0507094].
  
\bibitem{Laporta:2001dd}
  S.~Laporta,
  ``High precision calculation of multiloop Feynman integrals by difference equations,''
  Int.\ J.\ Mod.\ Phys.\ A {\bf 15} (2000) 5087
  [hep-ph/0102033].

\bibitem{Anastasiou:2004vj}
  C.~Anastasiou and A.~Lazopoulos,
  ``Automatic integral reduction for higher order perturbative calculations,''
  JHEP {\bf 0407} (2004) 046
  [hep-ph/0404258].

\bibitem{Kotikov:1990kg}
  A.~V.~Kotikov,
  ``Differential equations method: New technique for massive Feynman diagrams calculation,''
  Phys.\ Lett.\ B {\bf 254} (1991) 158.

\bibitem{Gehrmann:1999as}
  T.~Gehrmann and E.~Remiddi,
  ``Differential equations for two loop four point functions,''
  Nucl.\ Phys.\ B {\bf 580} (2000) 485
  [hep-ph/9912329].
  
\bibitem{Remiddi:1999ew}
  E.~Remiddi and J.~A.~M.~Vermaseren,
  ``Harmonic polylogarithms,''
  Int.\ J.\ Mod.\ Phys.\ A {\bf 15} (2000) 725
  [hep-ph/9905237].
  
\bibitem{Gehrmann:2001pz}
  T.~Gehrmann and E.~Remiddi,
  ``Numerical evaluation of harmonic polylogarithms,''
  Comput.\ Phys.\ Commun.\  {\bf 141} (2001) 296
  [hep-ph/0107173].
  
\bibitem{Vollinga:2004sn}
  J.~Vollinga and S.~Weinzierl,
  ``Numerical evaluation of multiple polylogarithms,''
  Comput.\ Phys.\ Commun.\  {\bf 167} (2005) 177
  [hep-ph/0410259].
  
\bibitem{Maitre:2005uu}
  D.~Maitre,
  ``HPL, a mathematica implementation of the harmonic polylogarithms,''
  Comput.\ Phys.\ Commun.\  {\bf 174} (2006) 222
  [hep-ph/0507152].
  
\bibitem{Maitre:2007kp}
  D.~Maitre,
  ``Extension of HPL to complex arguments,''
  Comput.\ Phys.\ Commun.\  {\bf 183} (2012) 846
  [hep-ph/0703052 [HEP-PH]].
 
  
\bibitem{Buehler:2011ev}
  S.~Buehler and C.~Duhr,
  ``CHAPLIN - Complex Harmonic Polylogarithms in Fortran,''
  arXiv:1106.5739 [hep-ph].
  
\bibitem{Anastasiou:2010pw}
  C.~Anastasiou, F.~Herzog and A.~Lazopoulos,
  ``On the factorization of overlapping singularities at NNLO,''
  JHEP {\bf 1103} (2011) 038
  [arXiv:1011.4867 [hep-ph]].
  
\bibitem{vanNeerven:1985xr}
  W.~L.~van Neerven,
  ``Dimensional Regularization Of Mass And Infrared Singularities In Two Loop On-shell Vertex Functions,''
  Nucl.\ Phys.\ B {\bf 268} (1986) 453.
  
\bibitem{Duhr:2011zq}
  C.~Duhr, H.~Gangl and J.~R.~Rhodes,
  ``From polygons and symbols to polylogarithmic functions,''
  arXiv:1110.0458 [math-ph].
  
\bibitem{Buehler:2012cu}
  S.~Buehler, F.~Herzog, A.~Lazopoulos and R.~Mueller,
  ``The Fully differential hadronic production of a Higgs boson via bottom quark fusion at NNLO,''
  JHEP {\bf 1207} (2012) 115
  [arXiv:1204.4415 [hep-ph]].
  
\bibitem{Gehrmann:2010ue}
  T.~Gehrmann, E.~W.~N.~Glover, T.~Huber, N.~Ikizlerli and C.~Studerus,
  ``Calculation of the quark and gluon form factors to three loops in QCD,''
  JHEP {\bf 1006} (2010) 094
  [arXiv:1004.3653 [hep-ph]].
  
\bibitem{Lee:2010cga}
  R.~N.~Lee, A.~V.~Smirnov and V.~A.~Smirnov,
  ``Analytic Results for Massless Three-Loop Form Factors,''
  JHEP {\bf 1004} (2010) 020
  [arXiv:1001.2887 [hep-ph]].
  
  \bibitem{axodraw}
  \verb+http://www.nikhef.nl/~form/maindir/others/axodraw/axodraw.html+
 
 

\end{thebibliography}

\providecommand{\href}[2]{#2}\begingroup\raggedright
\endgroup

\end{document}